\documentclass[useAMS,UKenglish]{mn2e}
\usepackage{times}
\usepackage{graphicx,lscape}
\usepackage{fix2col}
\usepackage{graphicx}
\usepackage{ulem}
\usepackage{rotating}
\usepackage{lscape}
\usepackage{longtable}

\def\ltsima{$\; \buildrel < \over \sim \;$}
\def\simlt{\lower.5ex\hbox{\ltsima}}
\def\gtsima{$\; \buildrel > \over \sim \;$}
\def\simgt{\lower.5ex\hbox{\gtsima}}

\def\kms{{$\; \rm km~sec^{-1}\;$}}

\def\vhb{{$\;V_{HB}\;$}}
\def\vvhb{{$\;V - V_{HB}\;$}}

\def\sumw{{$\;\rm \sum{W}\;$}}
\def\sigw{{$\;\rm \sigma_{\sum{W}}\;$}}
\def\wp{{$\;\rm W^\prime\;$}}

\def\sigwp{{$\;\rm \sigma_{W^\prime}\;$}}

\def\feh{{\rm[Fe/H]\ }}

\def\nodata{{---}}

\begin{document}

\newlength{\figwidth}
\setlength{\figwidth}{6.in}
\newlength{\fighalfwidth}
\setlength{\fighalfwidth}{2.97in}

\newcommand{\etal}{et al.\ }
\newcommand{\rAA}{{\AA \enskip}}

\title[Chemical Abundances in Leo~I \& II dSphs] {The Chemical Abundances of the Stellar Populations in the \\
Leo~I and Leo~II dSph Galaxies\thanks{Data presented herein were
obtained at the W.M.~Keck Observatory, which is operated as a
scientific partnership among the California Institute of
Technology, the University of California and the National
Aeronautics and Space Administration. The Observatory was made
possible by the generous financial support of the W.M.~Keck
Foundation.}}

\author[T. L. Bosler, T. A. Smecker-Hane, P. B. Stetson]
{Tammy L. Bosler,$^1$ Tammy
A.~Smecker-Hane$^2$, Peter B.~Stetson$^3$\\
$^1$National Science Foundation, Division of Astronomical Sciences.\\
$^2$Department of Physics \& Astronomy, 4129 Frederick Reines
Hall, University of California, Irvine, CA
92697--4575.\\
$^3$National Research Council of Canada,
    Herzberg Inst. of Astrophysics, Dominion Astrophysical
    Observatory, \\ \hspace{0.1cm}5071 West Saanich Road, Victoria, BC  V9E 2E7,
    Canada.}

\maketitle
\begin{abstract}
We have obtained calcium abundances and radial velocities for 102
red giant branch (RGB) stars in the Leo~I dwarf spheroidal galaxy
(dSph) and 74 RGB stars in the Leo~II dSph using the Low Resolution
Spectrograph (LRIS) on the Keck~I 10-meter Telescope.  We report on
the calcium abundances [Ca/H] derived from the strengths of the
Ca~II triplet absorption lines at 8498, 8542, 8662~\AA~in the
stellar spectra using a new empirical Ca~II triplet calibration to
[Ca/H].  The two galaxies have different average [Ca/H] values of
$-1.34 \pm 0.02$ for Leo~I and $-1.65 \pm 0.02$ for Leo~II with
intrinsic abundance dispersions of 1.2 and 1.0~dex, respectively.
The typical random and total errors in derived abundances are 0.10
and 0.17~dex per star. For comparison to existing literature, we
also converted our Ca~II measurements to [Fe/H] on the scale of
Carretta and Gratton (1997) though we discuss why this may not be
the best determinant of metallicity; Leo~I has a mean [Fe/H] $\rm =
-1.34$ and Leo~II has a mean [Fe/H] $\rm = -1.59$. The metallicity
distribution function of Leo~I is approximately Gaussian in shape
with an excess at the metal-rich end, while that of Leo~II shows an
abrupt cutoff at the metal-rich end. The lower mean metallicity of
Leo~II is consistent with the fact that it has a lower luminosity,
hence lower total mass, than Leo~I; thus the evolution of Leo II may
have been affected more by mass lost in galactic winds. Our direct
and independent measurement of the metallicity distributions in
these dSph will allow a more accurate star-formation histories to be
derived from future analysis of their CMDs.
\end{abstract}


\section{Introduction}

The goal of this project is to map the chemical abundance
distribution of the stellar populations of the Leo~I and Leo~II
dwarf spheroidal (dSph) galaxies in order to constrain the
physical processes that regulate their evolution.
Table~\ref{table-dsphs} lists the properties of each galaxy for
reference.

Dwarf spheroidal galaxies (dSphs) are low luminosity ($M_V \le
-14$ mag), low surface brightness ($\Sigma_V \simlt 22$
mag/arcsec$^2$), and low mass ($M_{tot} \simlt 10^7 {\rm
M}_\odot$) systems.  They have little or no interstellar gas
($M_{HI} \simlt 10^5 {\rm M}_\odot$), and are dark-matter
dominated.  For recent reviews, see Mateo (1998), Smecker-Hane \&
McWilliam (1999), and Grebel (2000).  Their stellar velocity
dispersions are typically $\sim 10$ km/sec, and therefore galactic
winds from the first generation of supernovae are expected to
efficiently rid them of gas and prevent extended or multiple
epochs of star formation.  Although they seem to lack the mass and
material to sustain star formation for a significant fraction of
the Hubble time, the CMDs of the dSph satellites of the Milky Way
have shown that most dSphs have {\it very complex} star formation
histories (SFHs) and chemical evolution. Cold dark matter
hierarchical galaxy formation models have difficulty explaining
extended epochs of star formation in dSphs. They are predicted to
begin forming stars at early epochs, but lose a significant amount
of gas due to photoevaporation during reionization, which curtails
their star formation activity (Barkana \& Loeb 1999). Klypin et
al.~(1999) also points out that CDM models vastly over predict the
number of low mass halos; the Milky Way and M31 have only
one-tenth the number of low mass satellites. Are these ``missing"
satellites dark because they never accreted gas, or were the dark
matter halos stripped of their baryons through supernovae-driven
winds? Determining the age of the oldest stars in the outer halo
satellites such as the Leo~I and Leo~II dSphs, whose
Galactocentric distances are 270 and 204~kpc, respectively, is
important because CDM models predict that these should form later
than the Milk Way's inner halo globular clusters and dSphs.

Almost every dSph in the Local Group appears to have had a unique
SFH.  From analysis of a CMD obtained with WFPC2 aboard the HST,
Mighell \& Rich (1996) have shown that the Leo~II dSph formed
essentially all of its stars from $\sim 7$ to 14~Gyr ago with 50\%
of its stars formed in a 4~Gyr period. In striking contrast, most
of the stars in the Leo~I dSph are younger than this; 87\% of its
star formation occurred from 1 to 7~Gyr ago and only 12\% occurred
$\simgt 10$~Gyr ago (Gallart et al.~1999a and 1999b).  More recent
photometry from Held \etal (2000), which includes a much larger
radial coverage than the HST data, also shows an extended blue
horizontal-branch in Leo~I indicating the existence of an old,
metal-poor population.  Held \etal (2001) also report the
discovery of candidate RR~Lyrae stars in the system, further
strengthening the case for an old population in Leo~I.

Defining the chemical evolution of the dSphs will be key to
understanding their complex SFHs because the chemical abundances
constrain the inflow and outflow of gas from the systems. Did some
dSphs accrete fresh gas that fuelled subsequent star formation?
Were outflows driven by supernovae so extensive that today's dSphs
have only a small fraction of their former mass? The chemical
abundances of dSph stars -- the mean metallicity as well as the
detailed distribution in metallicity  -- can be determined from
spectroscopy of individual red giant stars in the dSphs. Measuring
the metallicity distribution of individual stars will greatly
improve the SFHs derived from analysis of CMDs.

In order to understand the true spread in stellar chemical
abundances more precise abundance indicators than can be inferred
from CMD analysis are needed.  We, therefore, measured the abundance
distribution of the Leo~I and Leo~II dSphs from spectroscopy of
individual red giant stars using the Ca~II absorption lines in the
near infrared (8498, 8542, and 8662~\AA). We utilized the new
calibration of Ca~II to [Ca/H] from Bosler (2004, hereafter B04) to
derive abundances with random and total errors of 0.10 and 0.17~dex,
respectively.  This new empirical calibration to calcium abundances
avoids the inherent dependance on [Ca/Fe] built into the existing
Ca~II calibration to [Fe/H].  However, in order to allow for easier
comparison to earlier Ca~II work on Local dwarf galaxies, we also
calculated the [Fe/H] values of the stars based on the Rutledge
\etal (1999) calibration; however, we discuss the pitfalls of using
the Ca~II triplet to determine [Fe/H]. The new Ca~II calibration to
[Ca/H] is completely independent of star to star variations in
[Ca/Fe] unlike previously utilized Ca~II abundance calibrations.

\section{The Sample}

The photometry and astrometry for Leo~I and Leo~II were obtained
as part of a larger program to provide homogeneous photometry for
star clusters and nearby resolved galaxies (e.g., Stetson, Hesser
\& Smecker-Hane 1998, Stetson 2000). The present photometry is
based upon a large database of CCD images acquired by ourselves
and collaborators, images obtained from the archives of the
Canada-France-Hawaii Telescope and the Isaac Newton Group, and
images privately donated by other astronomers. In the present
case, the Leo~I photometric database contains some 30,000
different stars recorded in 449 individual CCD images taken in the
UBVRI filters (4 in U, 96 in B, 257 in V, 72 in R, and 20 in I)
covering an irregular area spanning an extreme range of
40.7$^\prime$ in the east-west direction by 40.6$^\prime$
north-south. For Leo~II, some 17,000 stars were measured in 124
CCD images taken in UBVRI filters (2 in U, 30 in B, 81 in V, 5 in
R, and 6 in I) spanning an area with extreme dimensions
16.9$^\prime$ by 16.5$^\prime$.  All observations were transformed
to the UBVRI photometric system of Landolt (1992; see also Stetson
2000) based on observations of primary (Landolt) and secondary
(Stetson) standard stars obtained during the same nights with the
same equipment.

The astrometry for the Leo~I and II photometric catalogs was based
on positions of reference stars taken from the United States Naval
Observatory Guide Star Catalogue~A2.0; (henceforth USNOGSC, Monet
et al.~1998), access to which was obtained through the services of
the Canadian Astronomy Data Center. The authors of the USNOGSC
claim a typical astrometric error of 0.25$\,$arcsec, which they
believe is dominated by systematic errors in the calibration
procedure.  When transforming relative $(x,y)$ positions from
large-format CCD images to absolute right ascensions and
declinations from the USNOGSC for stars in common, the typical
root-mean-square difference is 0.3 to 0.4$\,$arcsec in each
coordinate.  Some of this is clearly due to proper-motion
displacements accumulated during the forty-plus years between the
obtaining of the first Palomar Observatory Sky Survey and these
CCD data. However, a significant part of the differences is also
due to the lower angular resolution of the Schmidt plates as
scanned by the Precision Measuring Machine (built and operated by
the U.~S.~Naval Observatory Flagstaff):  particularly in crowded
fields such as these, a single entry in the USNOGSC is
occasionally found to correspond to the photocenter of a close
pair or a clump of several stars in the CCD imagery.

As a result of these non-Gaussian errors (i.e., proper motions and
blending), astrometric transformations were performed using an
iterative procedure wherein 20-constant cubic polynomials are used
to approximate the transformation of the $(x,y)$ positions
measured in the CCD images to standard coordinates obtained from
the USNOGSC (see Cohen, Briley, \& Stetson 2002 for more details).
When this procedure has been completed 706 detections common to
the USNOGSC were found and the data set in the field of Leo~I show
positional differences with a dispersion of 0.37 arcsec in both
the right ascension and declination directions; for Leo~II, 292
stars show positional agreement with a dispersion of 0.32 arcsec
in each direction.

The systematic errors of right ascensions and declinations on the
system of the USNOGSC are expected to be of order
$\sigma/\sqrt{N-20}\sim0.02\,$arcsec where $N$ is the number of
stars used in the astrometric solution. We can provide no
independent information on the accuracy with which the USNOGSC
coordinate system is referred to an absolute inertial reference
frame of right ascension and declination.  Individual {\it
random\/} errors in the coordinate measurements are probably not
much better than 0.02$\,$arcsec on a star-by-star basis, the
errors becoming somewhat worse where crowding is an issue.  The
logarithm of the observed flux in the stellar spectra is very well
correlated with our V-band magnitudes for the stars, and thus
there is no evidence of sizable relative position errors compared
to the size of the 1.0 arcsec-wide slits.  However while aligning
LRIS slit masks on the stars in a given field, one rotates the
field of view to get the best possible alignment, and thus the
flux measurements do not rule out possible systematic rotations of
the coordinate system with respect to the true inertial frame.
Still, these should not be present at levels much greater than the
0.02-arcsec precision claimed above.  A comprehensive catalog of
positions and photometry for these two galaxies is in preparation.

\begin{figure*}
     \includegraphics[width=\figwidth]{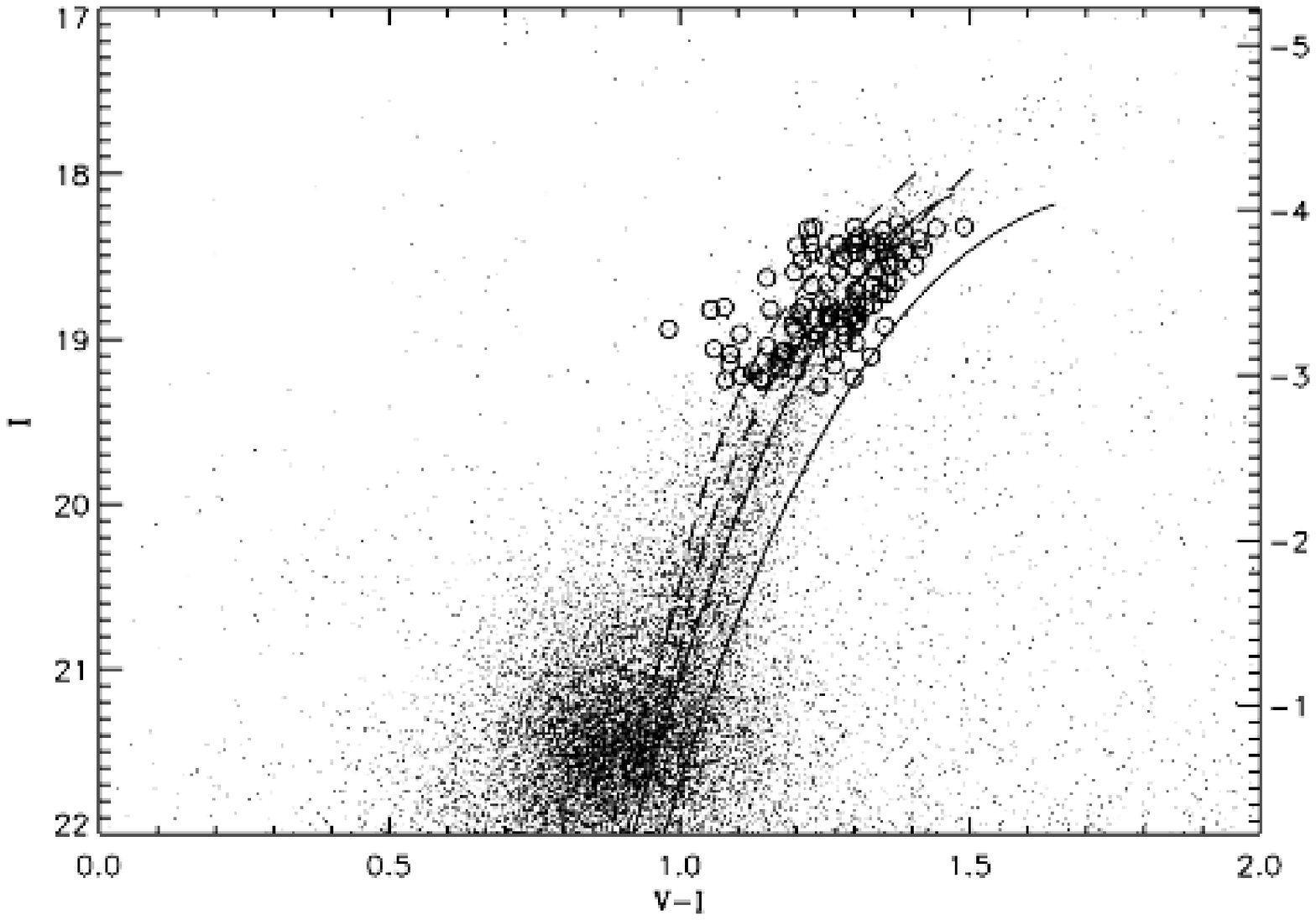}
   \caption{The color-magnitude diagram for the Leo~I dSph with
            the spectroscopic targets shown as large circles. Solid lines show
fiducials for Galactic globular clusters, from left to right, M15
with [Fe/H] $ = -2.0$ and M2 with \feh $=-1.31$ (Da Costa \&
Armandroff 1990).  The range in ages for the selected stars is
demonstrated with the dashed lines representing Yale isochrones
with [Fe/H] = $-$1.28 and with ages 2 and 4~Gyrs, from left to
right, respectively; stars bluer than the M2 fiducial are likely
younger. Note that, for convenience, e adopted the M15 fiducial to
describe the shape of the Leo~I RGB in selecting our spectroscopic
targets.}
            \label{fig-leo1CMD}
    \end{figure*}

\begin{figure*}
            \includegraphics[width=\figwidth]{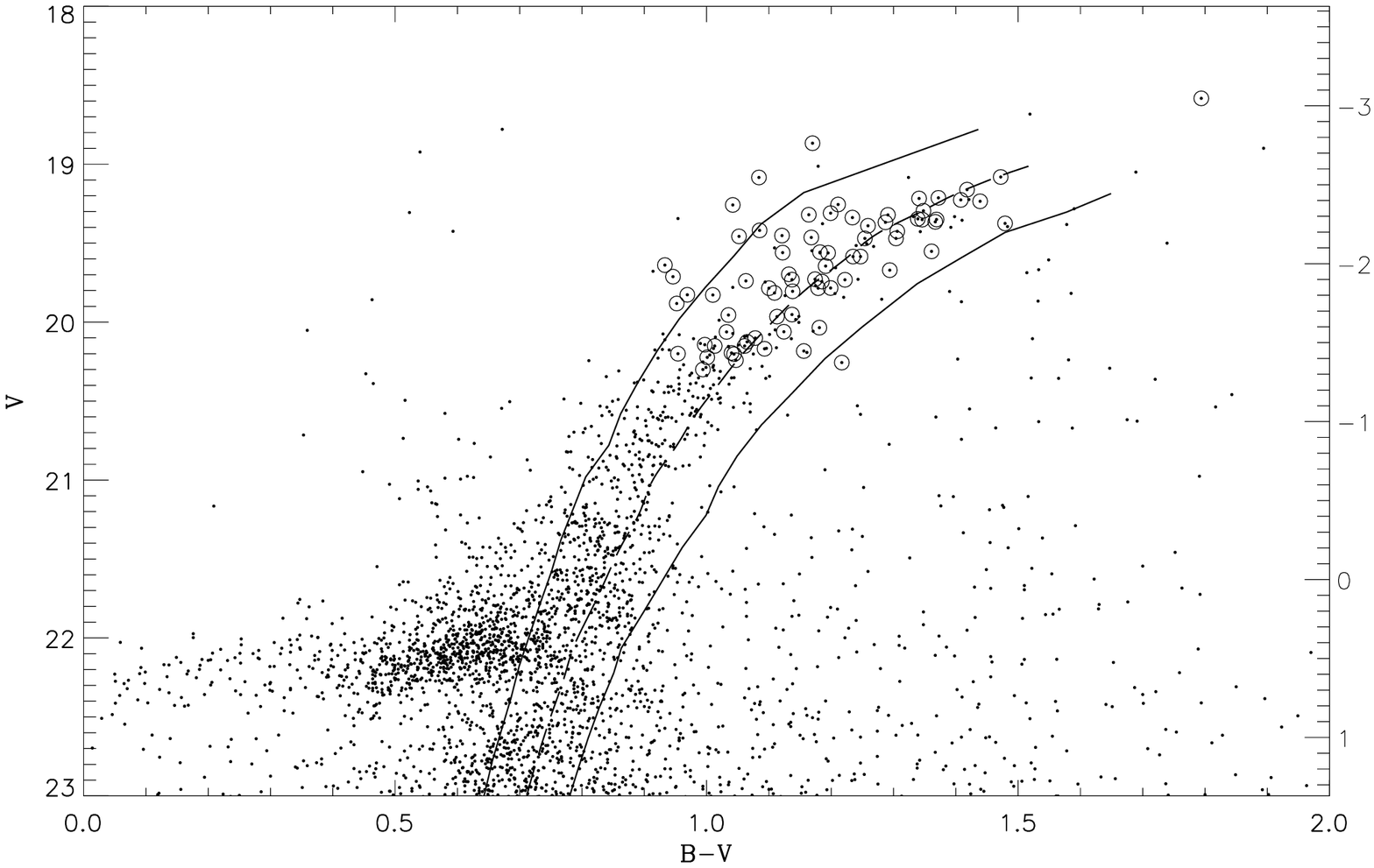}
            \caption{The color-magnitude diagram for the
Leo~II dSph with the spectroscopic targets shown as large circles.
Solid lines show fiducials for Galactic globular clusters, from
left to right, M68 with [Fe/H] $= -1.92$ (McClure et al.~1987) and
M5 with \feh $=-1.17$ (Sandquist et al.~1996). The dashed line
shows the adopted fiducial for the RGB of Leo II.}
\label{fig-leo2CMD}
    \end{figure*}

Our current CMDs for the Leo~I and Leo~II dSphs are shown in
Figures~\ref{fig-leo1CMD} and \ref{fig-leo2CMD}, respectively. The
stars in the spectroscopic sample are marked with large open
circles. On each CMD, fiducials for two Galactic globular clusters
with different metallicities are plotted for comparison. For
Leo~I, we have assumed a reddening of $\rm E(V-I) = 1.25 [ 1 +
0.06 (B-V)_0 + 0.014E(B-V)]\,\, E(B-V) = 0.054$ mag using the
reddening relationship derived by Dean, Warren \& Cousins (1978)
and assuming the mean color of the red giant branch (RGB) stars is
(B--V)$_0 = 1.2$. We have defined fiducial RGBs for each dSph,
which were used to quantify the color distribution in each dSph
and ensure that the spectroscopic targets span the color range,
and thus the selection is as representative in metallicity as
possible. For the Leo~I fiducial, we simply adopted the fiducial
for M15 = NGC~7078 (Da Costa \& Armandroff 1990), which has
[Fe/H]~$=-2.02$, because it lies roughly along the middle of the
Leo~I RGB. Because the of the age-metallicity degeneracy in the
CMD, the old (14~Gyr), metal-poor M15 RGB fiducial lies along the
same part of the CMD as the RGB of the younger (~4~Gyr),
intermediate-metallicity Leo~I RGB fiducial. Yale isochrones with
[Fe/H] = $-$1.28 and ages 2 and 4~Gyrs have been plotted on the
Leo~I CMD to illustrate the young ages of the stars on blue edge
of the RGB. For the Leo~II fiducial, we adopted the fiducial of
the Galactic globular cluster M68 = NGC~4590 (McClure, et
al.~1987), which has [Fe/H]~$= -1.92$, with an additional shift of
0.1 mag in (B--V) so that it ran along the middle of the Leo~II
RGB.

\begin{figure*}
      \includegraphics[width=\figwidth]{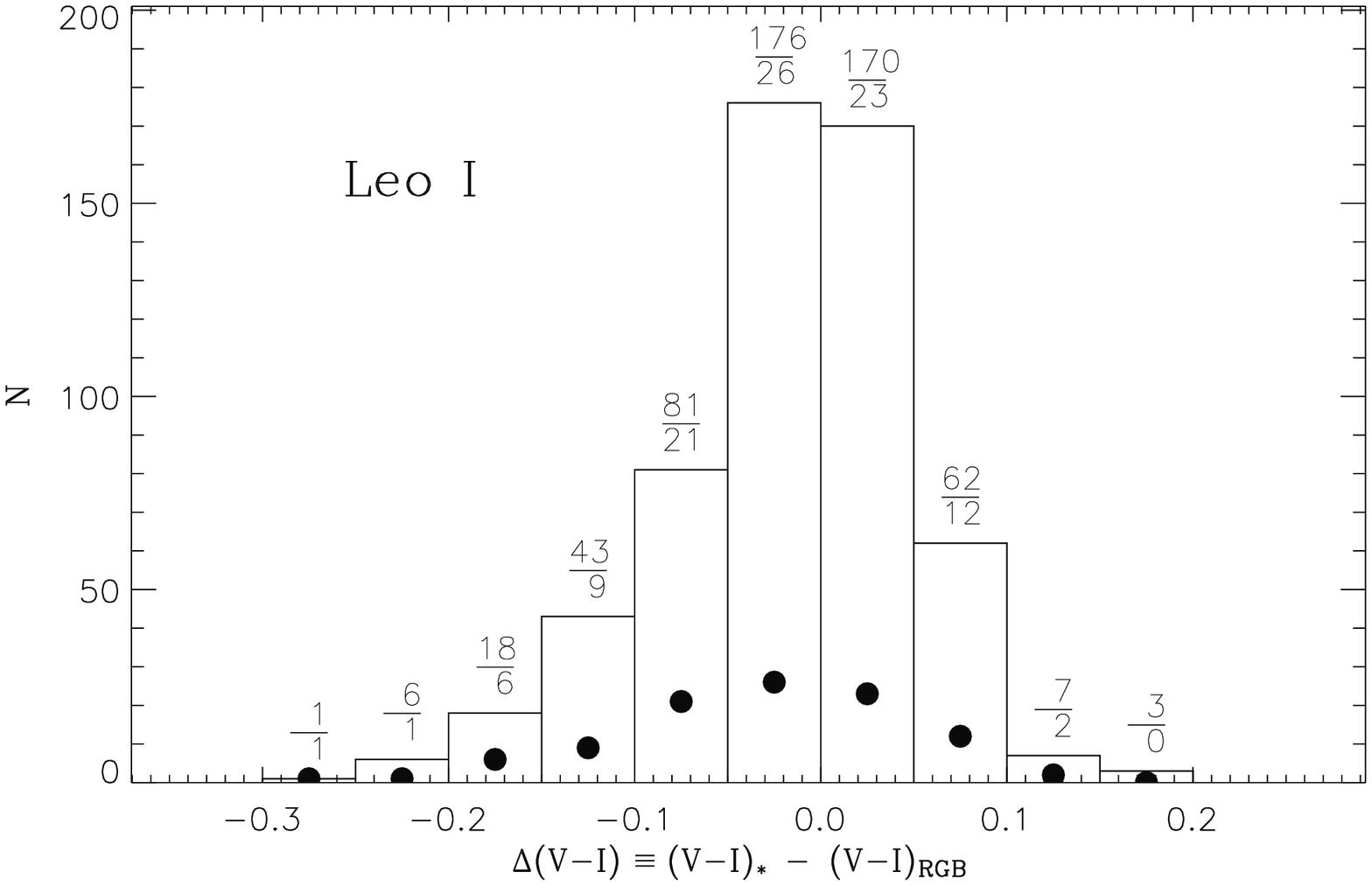}
          \caption{The histogram of the entire sample of RGB stars in the
Leo~I dSph (histogram) and the stars for which spectra were
obtained (filled circles). Above each bin is given the number of
stars in the bin in the total sample divided by the number of
stars in the bin in the spectroscopic sample.}
        \label{fig-select1}
      \end{figure*}

\begin{figure*}
    \includegraphics[width=\figwidth]{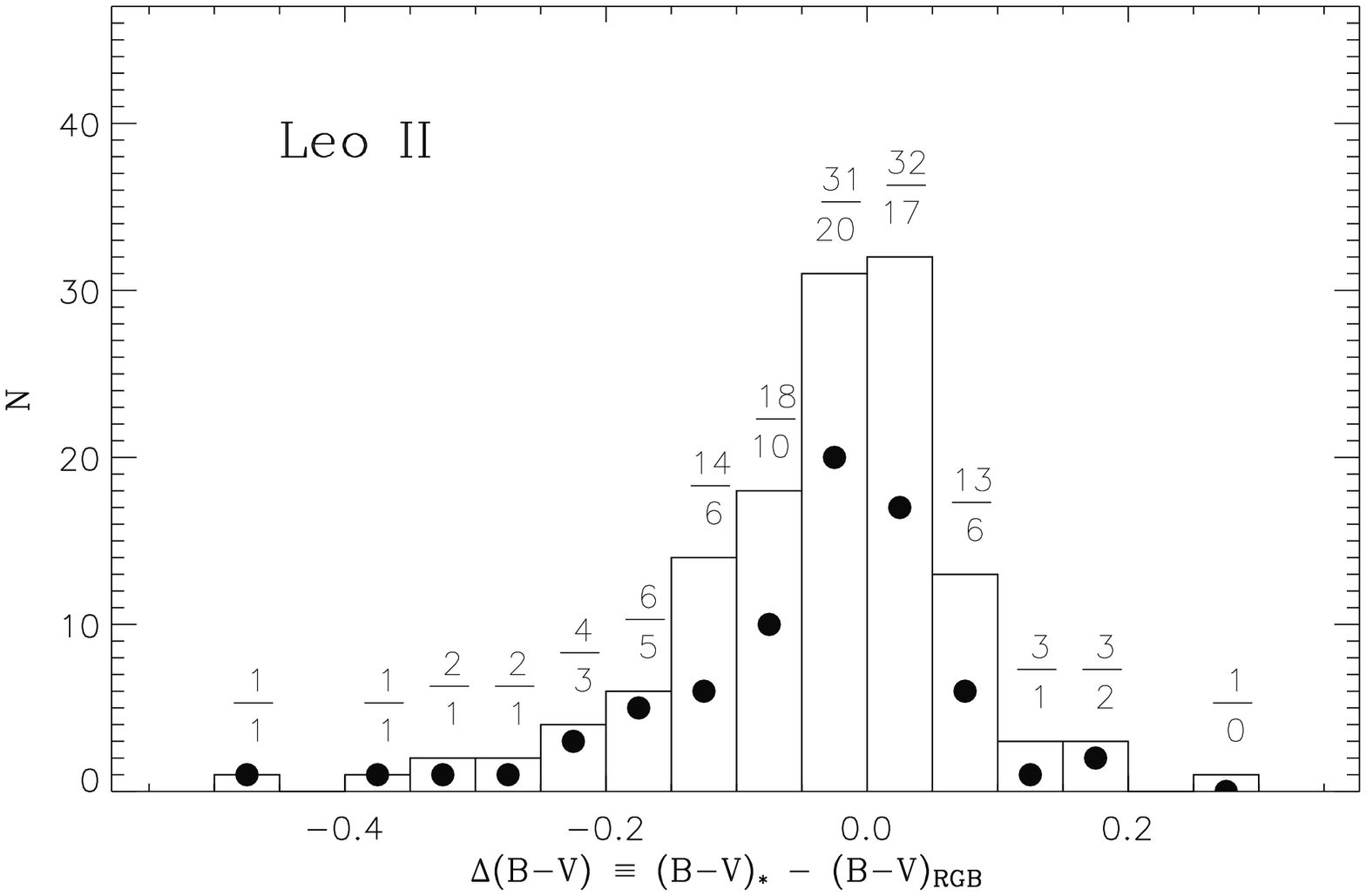}
     \caption{The histogram of the entire sample of RGB stars in the
Leo~II dSph (histogram) and the stars for which spectra were
obtained (filled circles). Above each bin is given the number of
stars in the bin in the total sample divided by the number of
stars in the bin in the spectroscopic sample.}
     \label{fig-select2}
     \end{figure*}

All dSph stars selected for observation have $\rm -3 \le V -
V_{HB} \le -2$ and $\rm 1.0 \le V - I \le 1.5$, where $\rm V_{HB}$
is the V-band magnitude of the horizontal branch. In order to have
the spectroscopic sample be as representative as possible with
respect to color and metallicity, we chose stars with the aim of
spanning the same distribution in color as the entire sample. To
quantify this, we defined the color difference, $\Delta_{\rm
(B-V)}$, to be the B--V color of an individual star minus the B--V
color of the fiducial RGB at the V-magnitude of the star. We used
B and V-band photometry to select Leo~II stars, and V and I-band
photometry to select Leo~I stars for spectroscopy. Note that the
images in the RI filters of the Leo~II fields were taken under
non-photometric conditions and remain uncalibrated. In designing
the slit mask configurations, we attempted to match the
distribution in $\Delta_{\rm (V-I)}$ and $\Delta_{\rm (B-V)}$ for
Leo~I and Leo~II, respectively. The histograms of color
differences are illustrated in Figures~\ref{fig-select1} and
\ref{fig-select2}.

Candidate red giants in the Leo~I and Leo~II dSphs were chosen for
spectroscopic observations based on their location in the CMD and
position in the slit mask field-of-view, which was $4^\prime
\times 7^\prime$ on the sky.  Using five slit mask orientations
for Leo~I we covered the core radius of the galaxy and 83\% of the
tidal radius along the east-west axis; only 56\% of the tidal
radius was covered along the north-south axis. Leo~II was observed
with four slit mask orientations, which covered an area of
$7^\prime \times 7^\prime$ on the sky; this covered all of the
core radius of the galaxy and approximately 80\% of the tidal
radius.

\section{Observations}

We obtained low-dispersion spectra of red giants in Leo~I and
Leo~II dSphs using the Keck~I 10-meter telescope and LRIS (Oke et
al.~1995) during 2 runs; run 1 was on 8 February 2002, and run 2
was on 5 March 2003 (UT). The pixel scale of the CCD in the
spatial direction was 0.215$^{\prime \prime}$/pixel. The CCD was
read out using two amplifiers with gains of 1.97 and 2.10
e$^-$/adu, and readnoises of 6.3 and 6.6 $\rm e^-$, for the left
and right amplifiers, respectively.  We used the 1200 $l$/mm
grating blazed at 7500~\AA, which gave a dispersion of
0.62~\AA/pix and resolution of 1.55 \AA\ (R $\approx$ 14,000). The
GG495 filter was used to block second-order light.  In multi-slit
mode, the available field of view was 4$^\prime \times 7^\prime$
for a minimum spectral coverage extending from 8355 -- 8890 \AA.
The slit widths were 1.0$^{\prime \prime}$~and the minimum slit
length was 6.0$^{\prime \prime}$. We were able to observe
approximately 20 to 30 stars per multi-slit mask.  In the
long-slit mode which was used for some of the calibration stars,
the slit was $1.0^{\prime \prime} \times 3.0^\prime$. The typical
seeing was 0.8$^{\prime \prime}$ as measured by the FWHM of the
profile of the spectra along the spatial direction.

Exposure times were $4 \times 900$ sec for most masks.
Observations of one field in Leo~I was only $\rm 2 \times 900$ sec
due to a pointing problem which caused the stars to be off center
of the slits for 2 exposures which were not used in the analysis.
Due to LRIS's sizable flexure, we obtained spectra of the Ne-Ar
arc lamp and a halogen flat after the final exposure on each
target for calibration purposes. In addition, a radial velocity
standard and telluric standards (rapidly rotating, B-type stars)
were observed in long-slit mode. Spectra of red giants in four
Galactic globular clusters were also taken using the slit maks
configuration to place our equivalent width measurements on the
well-defined scale of Rutledge et al.~(1997a and 1997b; hereafter
R97a and R97b).

\section{Data Reduction}

The usual CCD image reduction procedures of overscan fitting,
overscan subtraction, and zero subtraction were done using
routines we developed in the Interactive Data Language
(IDL)\footnote{IDL is commercial software sold by Research
Systems, Inc.}.  The two amplifiers have different zero levels,
which needed to be taken into account when creating the zero
image. Two bad pixels caused bleeding on the Amp~1 side and
created structure in the zero image.  Further reductions followed
those outlined in the data reduction tutorial on the LRIS web page
(http:$\backslash \backslash$alamoana.keck.hawaii.edu/inst/lris)
and Massey et al.~(1992) with slight modification in reduction
parameters. Cosmic ray removal, flat fielding in two dimensions,
extraction of spectra from two dimensions to one dimension, and
normalizing the continuum to unity were performed with tasks in
the {\sc IMRED} package of the Image Reduction and Analysis
Facility (IRAF)\footnote{IRAF is freely distributed by the
National Optical Astronomy Observatories, which is operated by the
Association of Universities for Research in Astronomy, Inc., under
cooperative agreement with the National Science Foundation.}.

Distortion in both the dispersion direction ($x$ axis) and spatial
direction ($y$ axis) had to be corrected in the 2-D spectra before
the spectra were extracted to 1-D.  We mapped and corrected
distortions along the $y$ axis of spectra using IRAF's {\sc
identify}, {\sc reidentify} and {\sc fitcoords} tasks on halogen
flats taken directly after each exposure. We applied the
distortion correction to the arc and object exposures using the
{\sc transform} task. To correct the distortion along the $x$
axis, we used the IRAF-based task {\sc xdistor} (Greg Wirth 2002,
private comm.) adapted to the specifications of the data set. This
task uses the strong night sky lines in the object spectra, or arc
lines in the Ne-Ar spectra, to map and correct the distortion in
the spectral direction for each aperture.  The distortion
correction was very accurate; after correction, the shifts in the
centroids of typical arc lines over the most distorted apertures
at the top and bottom of an image were $\simlt$ 0.05~pixels, which
is $\simlt$ 1.6\% of the FWHM.

Because the spectrograph has sizable flexure over the $\rm 4
\times 900$ sec observation times, the wavelength solution for
each aperture was determined using bright night sky lines in the
spectra themselves rather than the arc lamps taken after each set
of exposures.  Most of the night sky lines were blended at our
0.62~\AA/pix dispersion so we did not adopt the wavelengths of the
sky lines  determined from high-dispersion spectra (e.g.,
Osterbrock et al.~1996, Osterbrock \& Fulbright 1997).  Instead,
we calculated the central wavelength for approximately 40 bright,
night-sky emission lines in 25 apertures using Ne-Ar spectra taken
immediately after an exposure. Each stellar spectrum was
wavelength calibrated by using 25 to 40 night sky lines fit with a
spline3 (1st order) fit, which yielded wavelength solutions with
typical RMS residuals of 0.03~\AA. Each slit had a unique
wavelength coverage that fell somewhere in the range of 7770 --
9440 \AA.

Different spectra for individual targets were combined using the
IRAF {\sc scombine} task rejecting outliers at the $3\sigma$
level. The continua of all spectra were fit with cubic splines,
rejecting absorption features, and normalized to unity with the
{\sc continuum} task to produce the final calibrated spectra. We
measured the average signal-to-noise ratio per pixel (SNR) in each
spectrum by calculating the RMS in two wavelength windows where
absorption features are weak, 8580 -- 8620 \AA~and 8700 -- 8800
\AA, and averaged the results. The mean SNR was 18 for Leo~I stars
and 23 for Leo~II stars, and the range was $6 \leq$ SNR $\leq 42$
for all of the observed dSph stars. Abundances were only
determined for stars with SNR greater than 10 and for which at
least the two strongest Ca~II lines were observed.  These criteria
reduced the number of stars analyzed from 121 to 102 for Leo~I and
from 90 to 74 for Leo~II.

\section{Data Analysis}

In this section the measurement of heliocentric velocities, the
reduced equivalent widths of the Ca~II lines, and metallicities are
discussed.

\subsection{Heliocentric Velocities}
We verified membership of the stars in each dSph by deriving their
radial velocities because the intrinsic velocity dispersions of
these galaxies are well-defined, $\sim 10$ km/s, and very
different than stars in the Milky Way. We computed heliocentric
radial velocities, $v_{\rm helio}$, by cross-correlation (e.g,
Tonry \& Davis 1979) with a spectrum of IAU radial velocity
standard HD 12029 (K2III spectral type, B--V $=1.24$, $v_{\rm
helio} = 40.0$\kms) for Run~1 and HD 35410 (G9III spectral type,
$\rm B-V = 0.95$, $v_{\rm helio} = 20.3$ \kms) for Run~2.  We used
our IRAF task {\sc rviterate} to iterate the cross-correlation
performed by the IRAF {\sc fxcor} task.  {\sc rviterate} changes
the window of the observed wavelength in the object spectrum over
which the correlation is performed until the rest frame window of
8200 -- 9000 \AA~was used in both the template and object
spectrum. A ramp filter was used to filter out the large
wavelength fluctuations from any telluric bands on scales
$\approx$ 40~\AA, and small scale fluctuations from random noise.

\begin{figure*}
            \includegraphics[width=\figwidth]{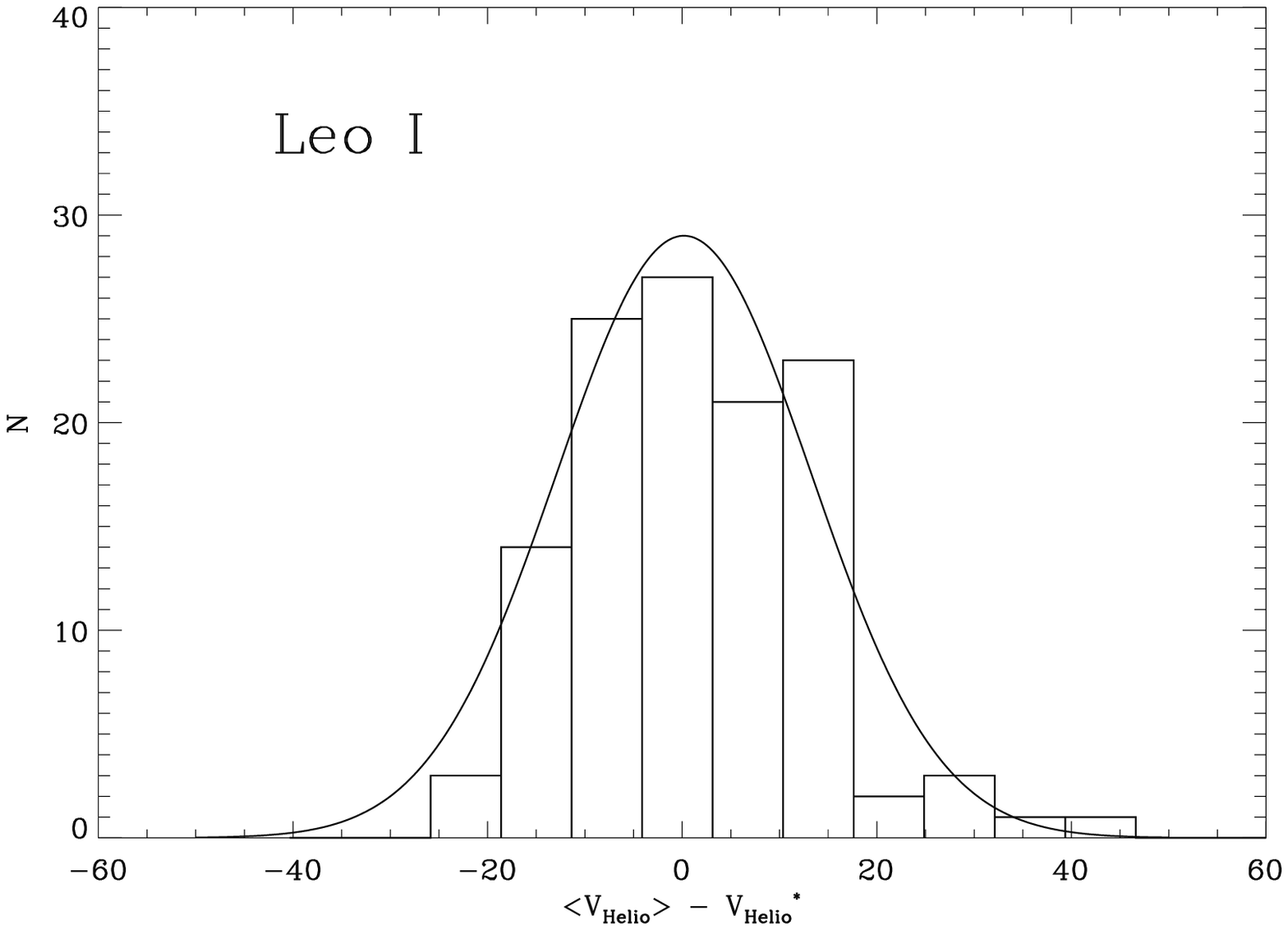}
            \includegraphics[width=\figwidth]{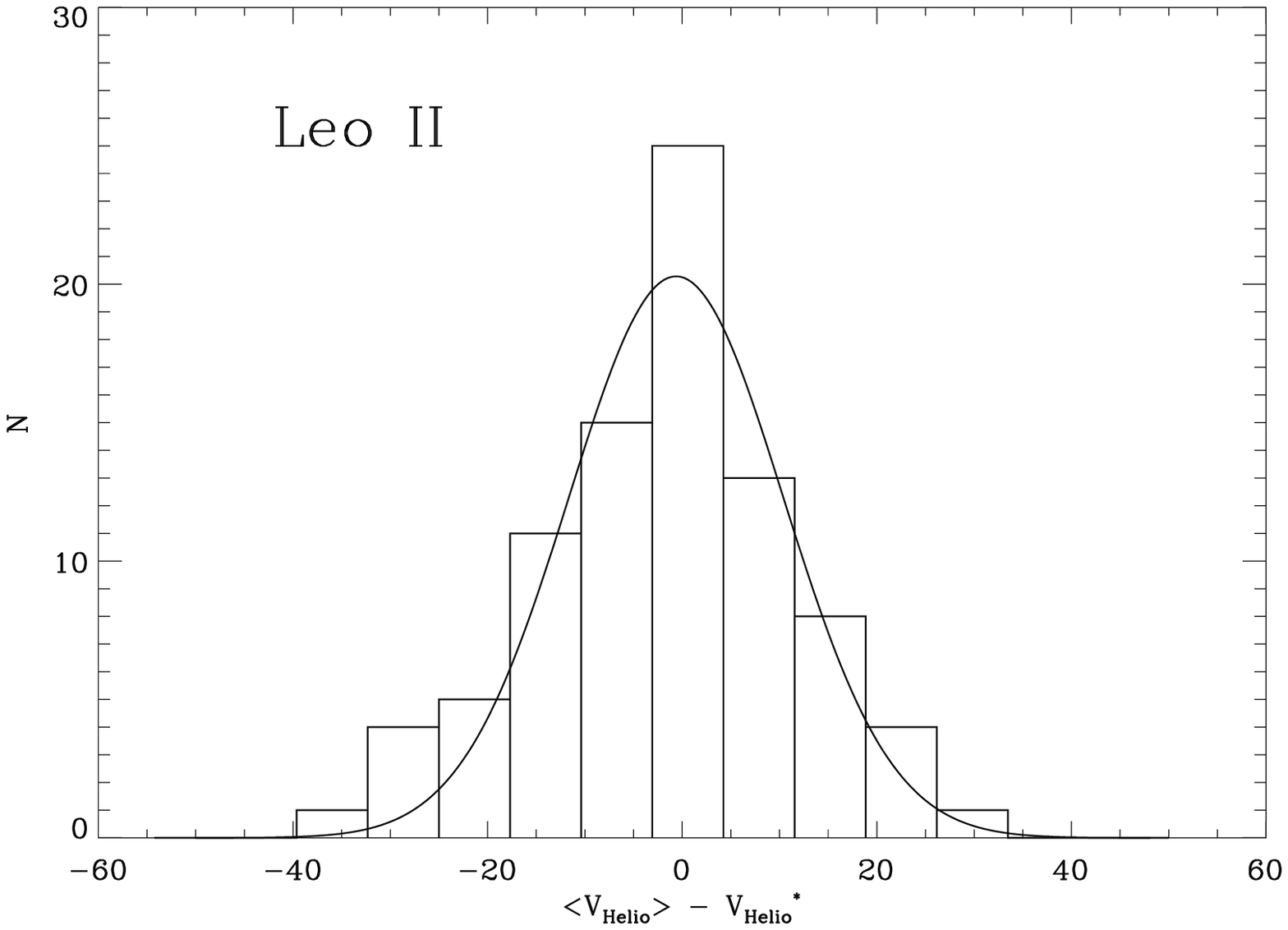}
            \caption{Velocity distributions of Leo~I and Leo~II overlaid with a Gaussian.}
            \label{fig-radvel}
            \end{figure*}

For 121 stars observed in Leo~I, the median velocity error is
9.2~$\rm km~sec^{-1}$. We found the average heliocentric velocity,
weighted by the errors, is $<v_{helio}> = +282.6~\pm~9.8~\rm
km~sec^{-1}$. Mateo, et al.~(1998) found $<v_{helio}> = +287.0
\pm~2.2$~\kms based on high-dispersion spectra of 15 RGB stars.
For 90 stars observed in Leo~II, the median velocity error is
8.5~$\rm km~sec^{-1}$. We found $<v_{helio}> = +85.8~\pm~8.4~\rm
km~sec^{-1}$. Vogt, et al.~(1995) found $<v_{helio}> = +76.0 \pm
1.3$~\kms and a dispersion of $\sigma_{v_{helio}}= 6.7 \pm
1.1$~\kms based on high-dispersion spectra of 31 red giants. The
differences between the mean velocities for this work and those of
earlier work are not significant given the differences in the
spectral resolution between the data sets. Our velocity resolution
is 54~\kms while that of Vogt \etal (1995) and Mateo, et
al.~(1998) is 9~$\rm km~sec^{-1}$.  For the case of Leo~I, the
great velocity dispersion (see Figure\ref{fig-radvel}) as well as
the small number of stars observed by Mateo \etal could also
contribute to the differences in derived average velocities.  As
we are more concerned with the deviations from the average rather
than the true velocities of the galaxies, we have used our mean
velocities for membership determination.

There can be systematic shifts in the radial velocities of stars
observed through multi-slit masks (e.g., Tolstoy et al. 2001) if
stars are not centered on the slits. Since our computed velocity
dispersions are similar to those determined from long slit work,
we can assume that our stars were typically well-centered on the slits.
There are no obvious correlations between a star's velocity
residual from the average $v_{helio}$ and its magnitude, color or radial
position within the galaxy.

Of the 121 stars in Leo~I, 90 have $v_{helio}$ within $3 \sigma$
of the average velocity, and all stars have $v_{helio}$ within $5
\sigma$. Of the 90 stars observed in Leo~II, 83 have $v_{helio}$
within $3 \sigma$ of the average velocity, and all stars have
$v_{helio}$ within $5 \sigma$. Figure~\ref{fig-radvel} illustrates
the velocity distributions for each galaxy; as no obvious velocity
outliers are apparent, we can assumed each star observed is a dSph
member.

\section{Chemical Abundance Analysis}

\subsection{The Reduced Equivalent Widths of the Ca~II Triplet}
\begin{figure*}
            \includegraphics[width=\figwidth]{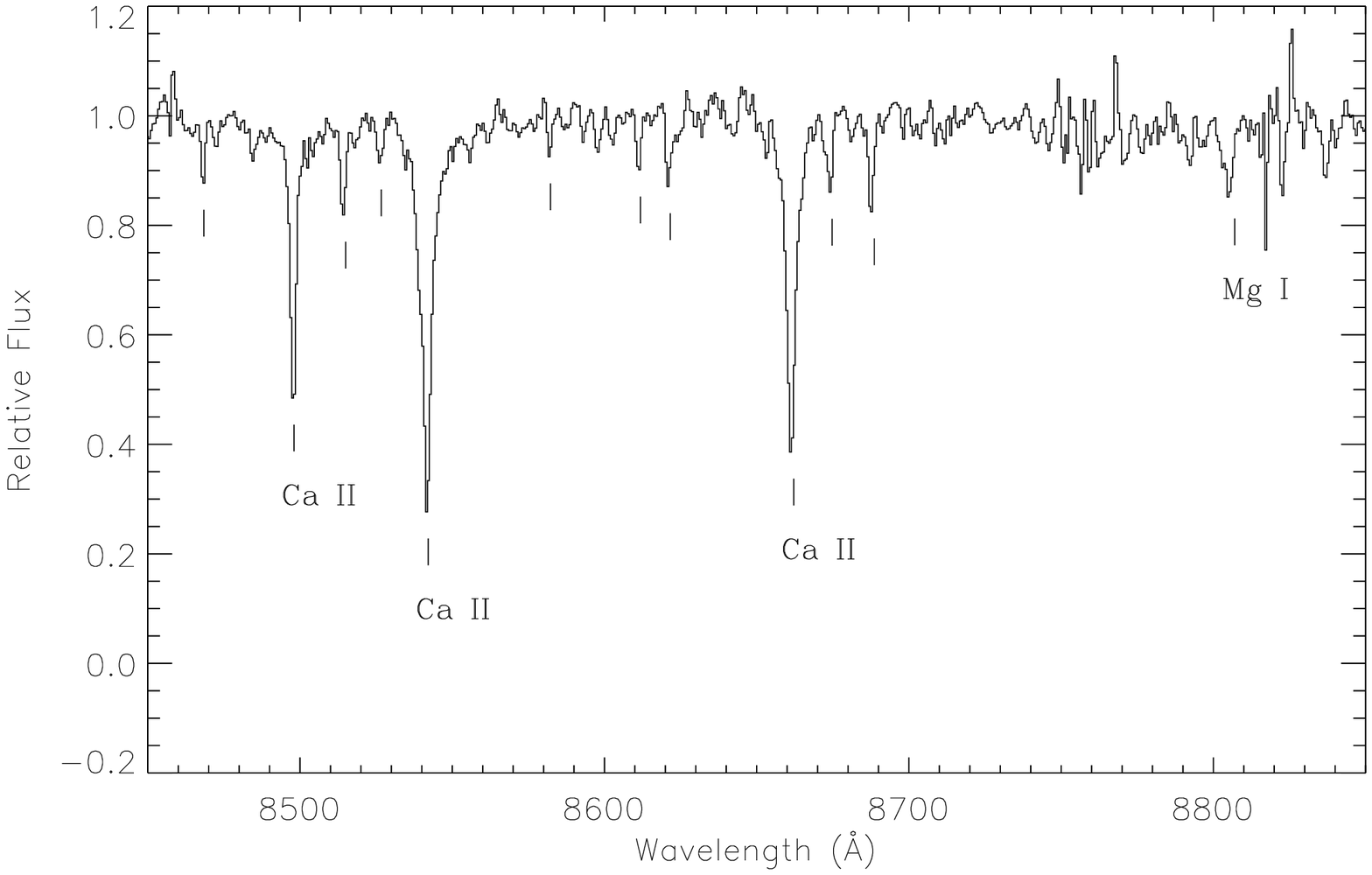}
            \caption{Low-dispersion spectrum of Star \#9915 in the Leo~I dSph
with $\rm [Ca/H] = -0.97$ and SNR = 35.  For clarity, Fe~I lines
are marked but not labelled.  Note that the Ca~II lines are
resolved but the weak Fe~I lines are not. Indeed many of the Fe I
lines are actually blends at this resolution (R=14,000). }
            \label{fig-leospec}
            \end{figure*}

\begin{figure*}
            \includegraphics[width=\figwidth]{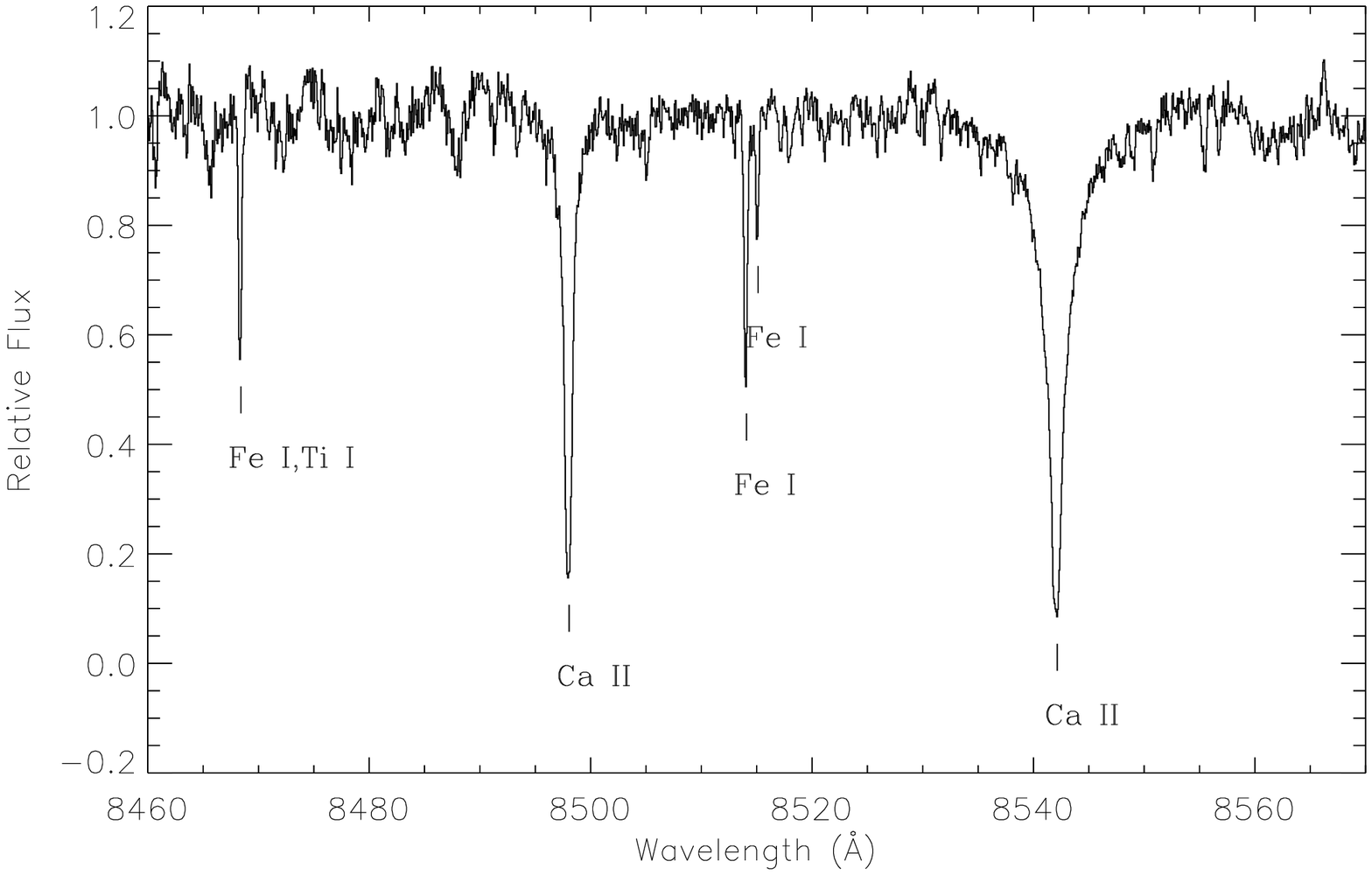}
            \caption{High-dispersion spectrum of calibration star HD135148,
which as $\rm [Fe/H] = -1.81$~and spectral type K0III (Bosler
2004). The wavelength coverage is much narrow than that of Fig
\ref{fig-leospec}. Note how the two Fe~I lines blended in the low
dispersion spectrum are resolved at higher resolution (R=40,000)
while the Ca~II lines are resolved at both resolutions (see
Figure~\ref{fig-leospec}).}
            \label{fig-HDspec}
            \end{figure*}

The Ca~II triplet absorption lines at 8498, 8542 and 8662~\AA~are
very strong in spectra of red giant stars, with equivalent widths
of order a few Angstroms. Figures \ref{fig-leospec} and
\ref{fig-HDspec} show how the strengths of the Ca~II lines compare
to strengths of the Fe~I lines in low and high-dispersion spectra.
The most widely used method of relating the equivalent widths of
the Ca~II triplet to metallicity is given by Rutledge \etal (1997a
and 199b; R97a and R97b, respectively).  R97a measured the reduced
equivalent widths, \wp (defined below) of the Ca~II triplet for
stars in 52 Globular clusters and, in R97b, found the empirical
relationship between \wp and [Fe/H] for the metallicity scales of
Carretta \& Gratton (1997, CG97) and Zinn \& West (1984, ZW84).

We have followed the R97a definition of sum of the equivalent
widths of the Ca~II lines,

\begin{equation}
\rm \sum{W} = 0.5~W_{8498} + W_{8542} + 0.6~W_{8662}.
\end{equation}

\noindent Lower weight is given to the weaker lines to yield
higher precision in the final index.  The pseudo-equivalent widths and errors of the
Ca~II lines were measured with an IRAF-based FORTRAN code named
{\sc EW} (G. Da Costa 1999, private comm.). The program assumes
the same line window regions used by R97a and the continuum
windows used by Armandroff \& Zinn~(1988).

\begin{figure*}
            \includegraphics[width=\figwidth]{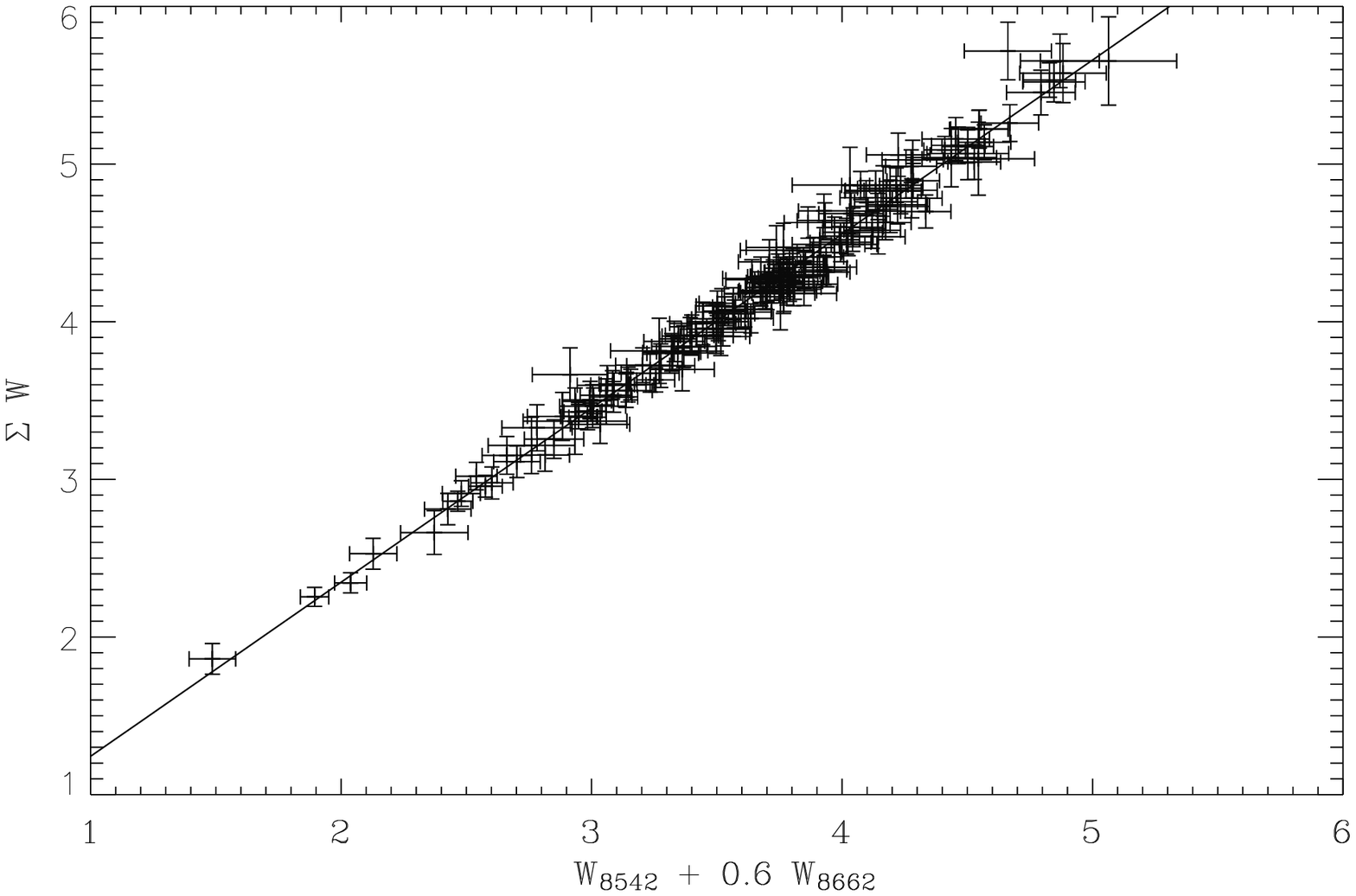}
            \caption{The relationship between the weighted sum of the two strongest Ca~II
            lines and the weighted sum of all three Ca II lines in
            the globular cluster stars.}
            \label{fig-wcalib}
        \end{figure*}

We were able to measure all three Ca~II lines in 155 of the
observed spectra, but in 20 spectra we could only accurately
measure the equivalent widths of the two strongest lines at
8542~\AA\ and 8662~\AA\ due to their lower SNRs.  In order to
obtain a measurement of \sumw for these stars, we determined the
relationship between \sumw and the equivalent widths of the two
strongest lines from the cluster stars with high SNR spectra. We
found the relationship was linear with

\begin{equation}
\rm \sum{W}  = {\it m} \,\, (W_{8542} + 0.6 \,\, W_{8662}) + {\it
b},
\end{equation}

\noindent where $m = 1.104 \pm 0.018$ and $b = 0.138 \pm
0.060$~\AA\ were the coefficients derived from least-squares
fitting analysis that incorporated errors on both values (The data
and fit are shown in Figure~\ref{fig-wcalib}).

The strengths of the Ca~II lines are most sensitive to stellar
gravity and abundance and much less sensitive to effective
temperature (e.g, D\'{i}az, Terlevich \& Terlevich 1989, Jorgensen
et al.~1992, and Cenarro et al.~2002). Using the ``reduced
equivalent width", $\rm W^\prime$, as defined by Armandroff \& Da
Costa (1991) allows the effect of gravity to be removed to first
order. The reduced equivalent width of a star is defined to be
\begin{equation}
\rm W^\prime = \sum{W} + 0.64 (V - V_{HB}),
\end{equation}

\noindent where $\rm V$ and $\rm V_{HB}$ are the V-band magnitudes
of the observed stars and the cluster or dSph horizontal branch,
respectively. R97a found a linear relationship existed between
$\rm (V - V_{HB})$ and $\rm \sum{W}$ for stars of an individual
globular cluster, and that the slope of 0.64 $\pm$ 0.02 \AA/mag
was insensitive to the metal abundance of the cluster.

In order to tie our \wp\ measurements to the well-calibrated and
often utilized scale of R97a, we observed 41 red giants in four
Galactic globular clusters. Table~\ref{table-clusters1} lists the
clusters, and gives the horizontal branch magnitude, the
metallicity, the average \wp~and error for the cluster from R97a,
and the \wp from this investigation. Values of \wp for NGC~1904
(M79), NGC~4590 (M68), and NGC~6171 (M107) are listed in R97a, but
they did not include NGC~5272 in their study. However, B04
observed stars in NGC~5272 in the Ca~II recalibration project at
the UCO Lick and W.M. Keck Observatories and placed its
measurements on the R97a scale. The [Fe/H] values for M79, M68 and
M3 listed in Table~\ref{table-clusters1} come from Bosler (2004;
see section \ref{abund_sec}) and were determined from atmospheric
abundance analysis of Fe~II lines in high-dispersion spectra. The
[Fe/H] value listed for M107 is from the R97b calibration of \wp
to [Fe/H] using the Carretta \& Gratton (1997, hereafter CG97)
metallicity scale. Table~\ref{table-clusters2} lists the
individual cluster stars and their measured equivalent widths in
this study and in that of R97a/R97b. The data on \sumw for
individual stars in the R97a study was provided by J. Hesser
(1999, private communication to T. Smecker-Hane). Note that
because of geometrical constraints inherent in using slit masks,
our sample of cluster stars is not identical to that observed by
R97a.

\begin{figure*}
            \includegraphics[width=\figwidth]{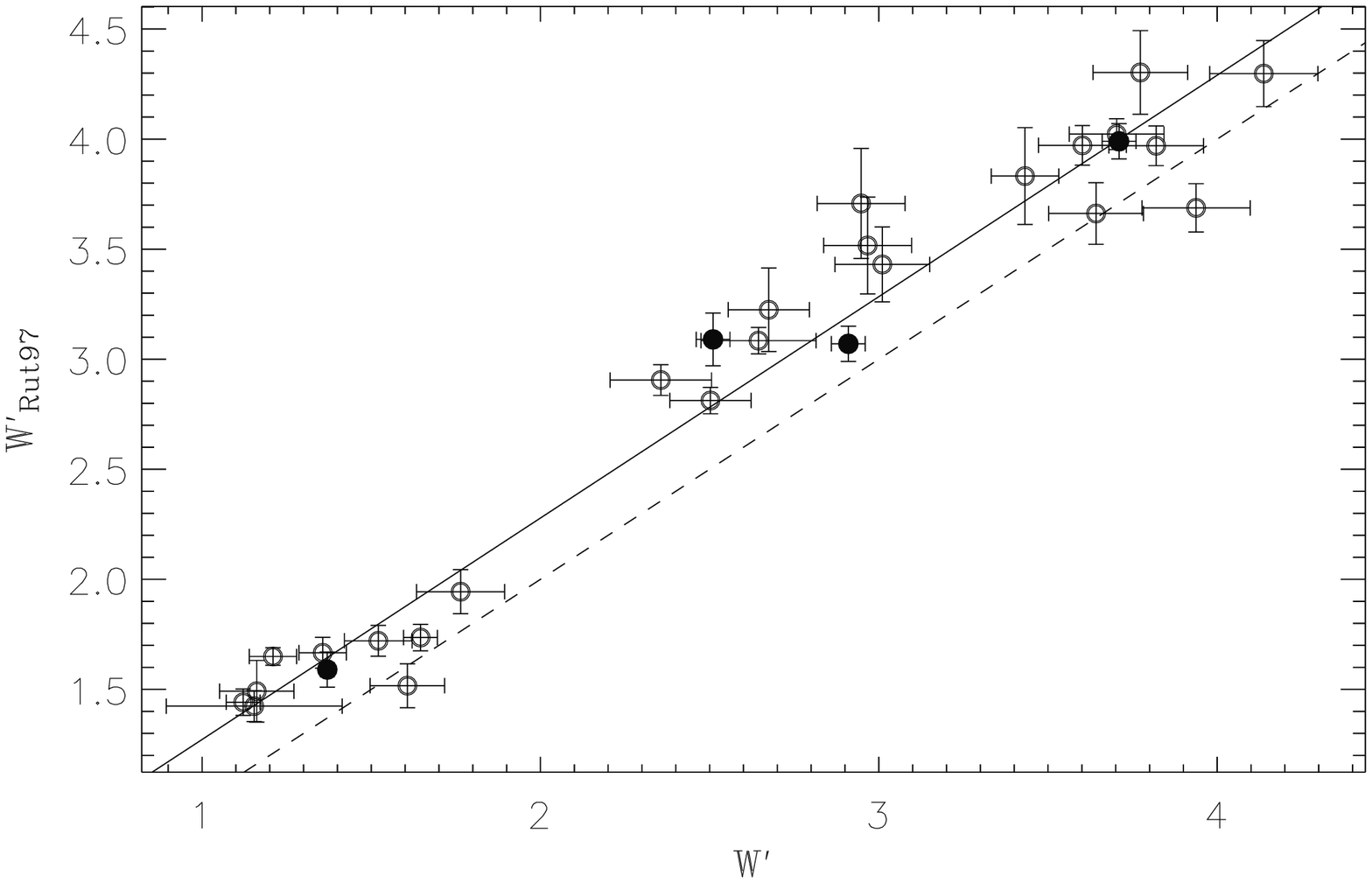}
            \caption{The reduced equivalent width measured by us and R97a
for individual globular cluster stars in NGC~1904, NGC~4590,
NGC~5272, and NGC~6171 (open circles), and the averages for the
clusters (filled circles). The dashed line shows the line of
equivalence. The solid line shows the regression using the \wp
values for individual stars.}
    \label{fig-wpscale}
    \end{figure*}

Comparing \wp~for the 23 individual cluster stars that the two
data sets have in common, we find

\begin{equation}
\rm W^\prime_{R97} = 1.006\,\, (\pm 0.025) \,\, W^\prime_{LRIS} +
0.267\,\,   (\pm 0.063),
\end{equation}

\noindent where $\rm W^\prime_{R97}$ is the value of \wp on the
scale of R97a and $W^\prime_{LRIS}$ is our observed value. The
least-squares fitting technique incorporated the errors in both
$\rm W^\prime_{LRIS}$ and $\rm W^\prime_{R97}$. The data and fits
are shown in Figure~\ref{fig-wpscale}. The small differences in
the equivalent widths measured by us and R97a are not surprising,
because the spectral resolution of our observations are
significantly different (our resolution is 1.55~\AA\ compared to
their resolution of 4 \AA), and the Ca~II lines are pressure
broadened and partially resolved at this resolution. In addition,
the equivalent widths are measured relative to the
pseudo-continuum (neither their spectra nor ours were flux
calibrated) and thus are slightly sensitive to the spectrograph's
efficiency. The differences we find are comparable to those R97a
find between their work and others (e.g., Table 6 in R97a).

The \wp values for Leo~I and Leo~II were determined using \vhb
values from the literature, and the final \wp values were places
on the scale of R97a to remain consistent with other literature on
the Ca~II triplet. For the Leo~I dSph, we assumed \vhb $= 22.60
\pm 0.12$ mag based on the RR Lyrae work of Held, et al.~(2001),
and for the Leo~II dSph, we assumed \vhb $= 22.17 \pm 0.14$ mag
based on the RR Lyrae work of Siegel \& Majewski~(2000). Note that
$\rm V_{HB}$ does change with the age of a stellar population, but
this change is small as long as the ages are $\simgt 3$~Gyr. This
is true for the Leo~II dSph, where ages are $\simgt 7$~Gyr
(Mighell \& Rich 1996), but the Leo~I dSph does have a few stars
as young as $\sim 2$~Gyr (Gallart et al.~1999b, Dolphin 2002). The
HB clump in Leo~I has a V magnitude that is $\sim 0.15$ magnitudes
brighter than the RR Lyrae stars. Thus adopting the RR Lyrae
magnitude will only result in a 0.05~dex systematic metallicity
error, in the sense that the true metallicity of a younger stars
would be more metal-rich than that inferred. This error is small
relative to the other errors, and hence is not taken into account
in the present analysis.

Note that although the Ca~II lines are easy to observe in low
resolution and low signal-to-noise spectra, abundances
\textit{cannot accurately be derived from modelling the the Ca~II
lines} due to the complexity of their formation. Additionally,
high dispersion spectra can resolve the weaker iron and neutral
calcium lines.  These lines can be accurately modelled to derive
stellar abundances, but these observations require long exposure
times and become impractical for surveying large numbers of faint,
distant stars. Thus, if one intends to use observations of the
Ca~II lines to determine stellar abundances, an empirical
relationship must exist between the equivalent widths of the Ca~II
lines are actual stellar abundances.

\subsection{Metallicity Calibration}\label{abund_sec}

Though the R97b calibration is widely used to derive abundances
from the Ca~II triplet, Rut97b points out that the two metallicity
scales for which calibrations where derived (CG97 and ZW84) can
yield dramatically different values of [Fe/H] for a given \wp.
Additionally, R97b points out that there is no reason to assume
that the relationship should be linear (as in the case of the CG97
metallicity scale) or non-linear (as in the case of the ZW84
metallicity scale). They discussed an apparent non-linearity
between the metallicity scales of ZW84 and also of Cohen
(compilation in Frogel \etal 1983), who examined more metal-rich
stars than were examined in other metallicity literature data.
Kraft \& Ivans (2004, hereafter KI04) also found a possible
non-linearity in the Ca~II to [Fe/H] relationship for metal-rich
stars in their investigation. In Cole \etal 2004, the effects of
line fitting on the equivalent widths of the Ca~II lines for
metal-rich stars also caused slight shifts in the empirical
relationship between Ca~II and [Fe/H]. One of the possibilities
noted in R97b for a non-linear relationship between \wp and [Fe/H]
for metal-rich stars is a changing [Ca/Fe] as a function of
[Fe/H].

\begin{figure*}
        \begin{center}
            \vspace*{0.1 in}
            \includegraphics[width=\figwidth]{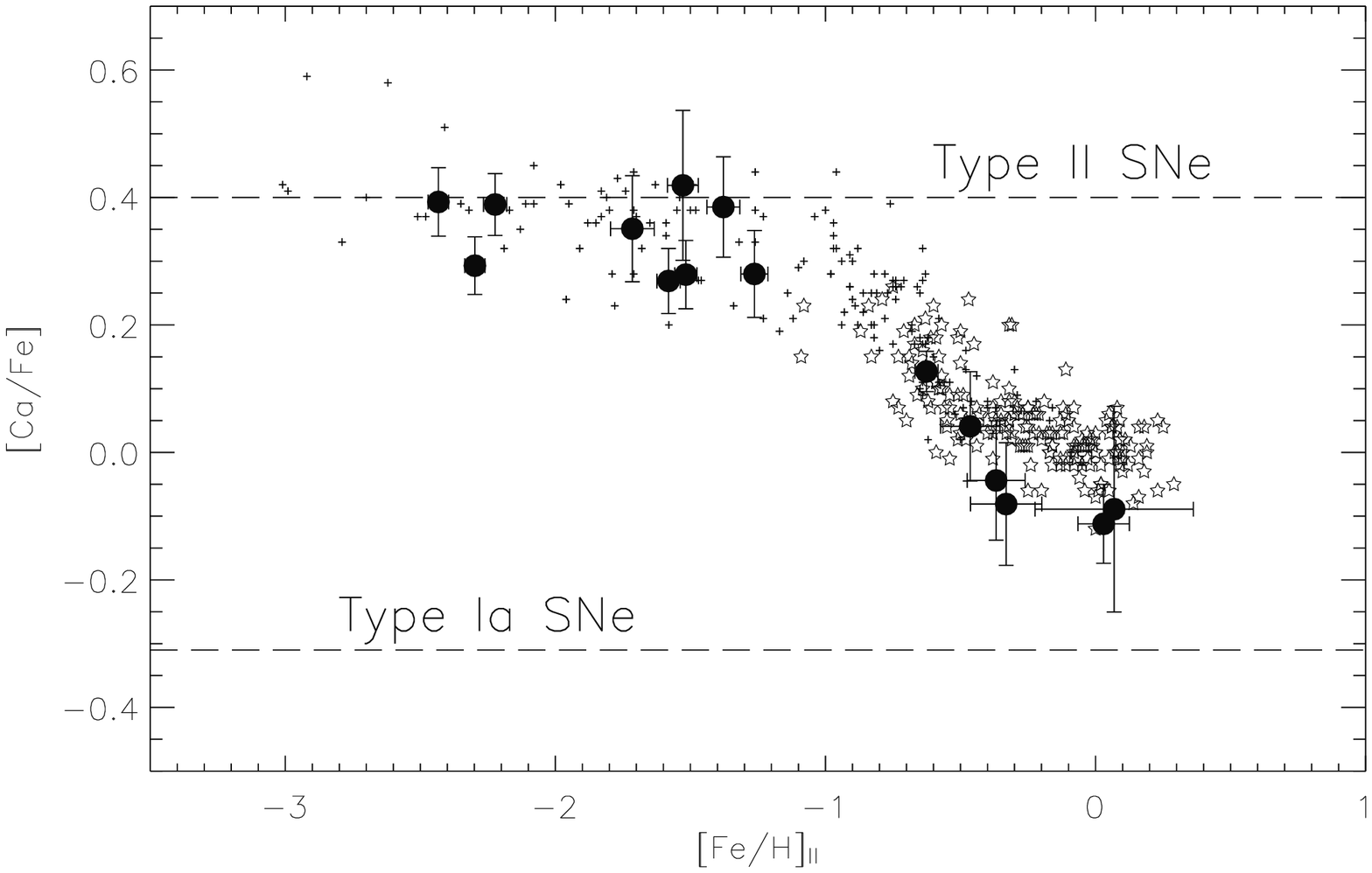}
            \caption[{Stellar $\rm [Ca/Fe]$ Vs. $\rm [Fe/H]$ }]{The
            [Ca/Fe] ratio as a function of [Fe/H] for all Galactic clusters from
            Bosler(2004; large filled circles) and
            individual Milky Way field stars from Fulbright (2000; pluses)
            and Edvardsson \etal (1993; open stars). Dashed horizontal lines
            indicate the predicted abundance ratios from Type~II and Type~Ia SNe
            ($+$0.4 from McWilliam, 1997
            and $-$0.31 from Thielemann \etal 1986, respectively).}
            \label{cafe_plot}
        \end{center}
    \end{figure*}

Figure~\ref{cafe_plot} plots [Ca/Fe] versus [Fe/H] for Galactic
clusters from B04 and Galactic field stars observed by Edvardsson
\etal (1993) and Fulbright (2000, 2002). The [Ca/Fe] in Milky Way
stars changes as a function of [Fe/H], and at [Fe/H] $\approx$
$-$0.8, the [Ca/Fe] value to change more rapidly as a function of
[Fe/H].

One major problem with determining a standard relationship between
Ca~II and [Fe/H] for stars with a range of ages and metallicities is
that the equivalent widths of the Ca~II lines are a function of,
among other things, [Ca/H] for RGB stars (see Jorgensen \etal 1992).
When one uses the equivalent widths of the Ca~II lines to infer
[Fe/H], one assumes some constant or smooth [Ca/Fe] relationship.
Applying the existing Ca~II to [Fe/H] calibration to extragalactic
stars is extremely dubious because one must assume that all observed
stars have a constant [Ca/Fe] value of approximately 0.3~dex (that
of the GGCs).

\begin{figure*}
        \begin{center}
            \includegraphics[width=\figwidth]{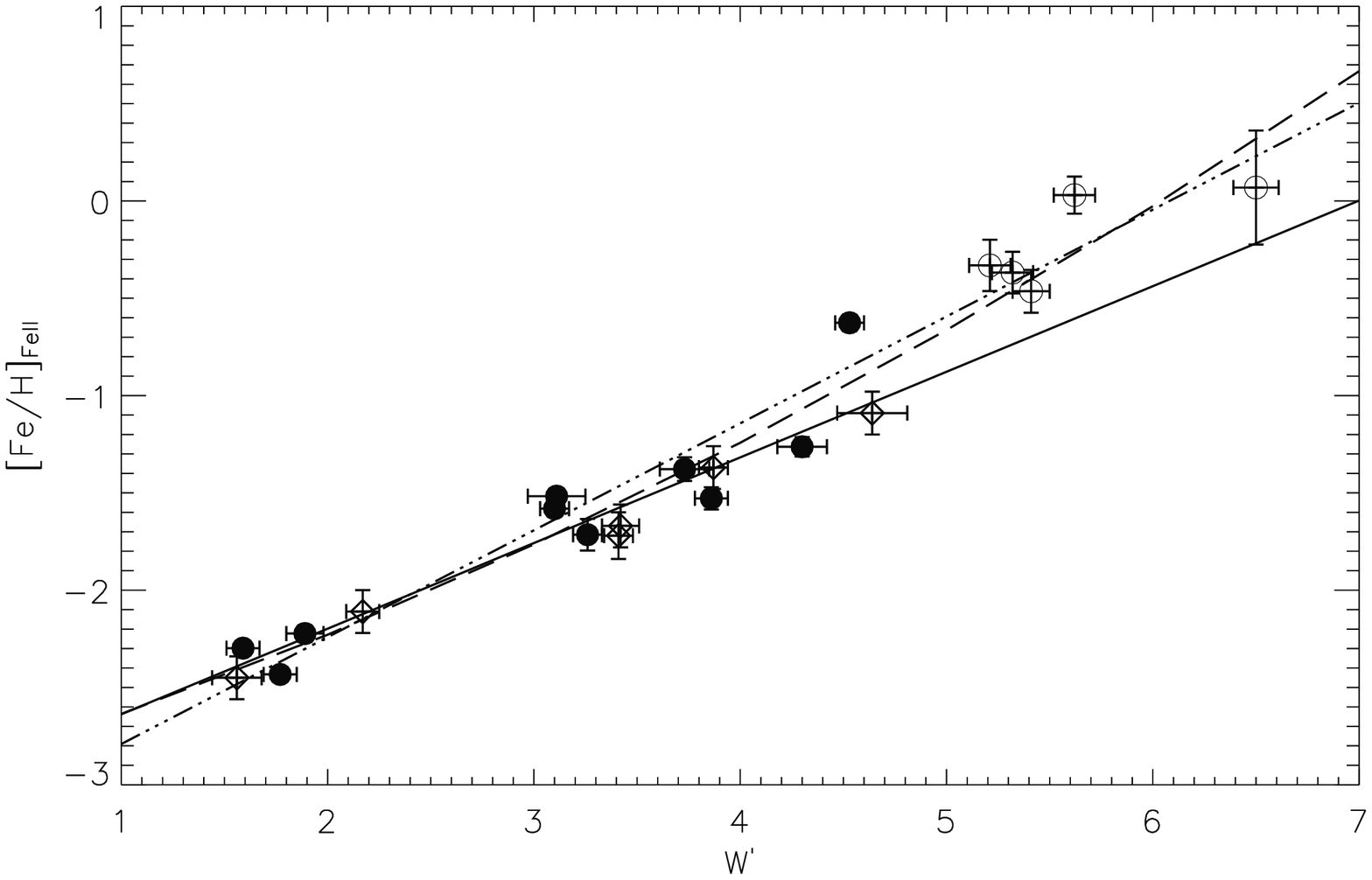}
            \includegraphics[width=\figwidth]{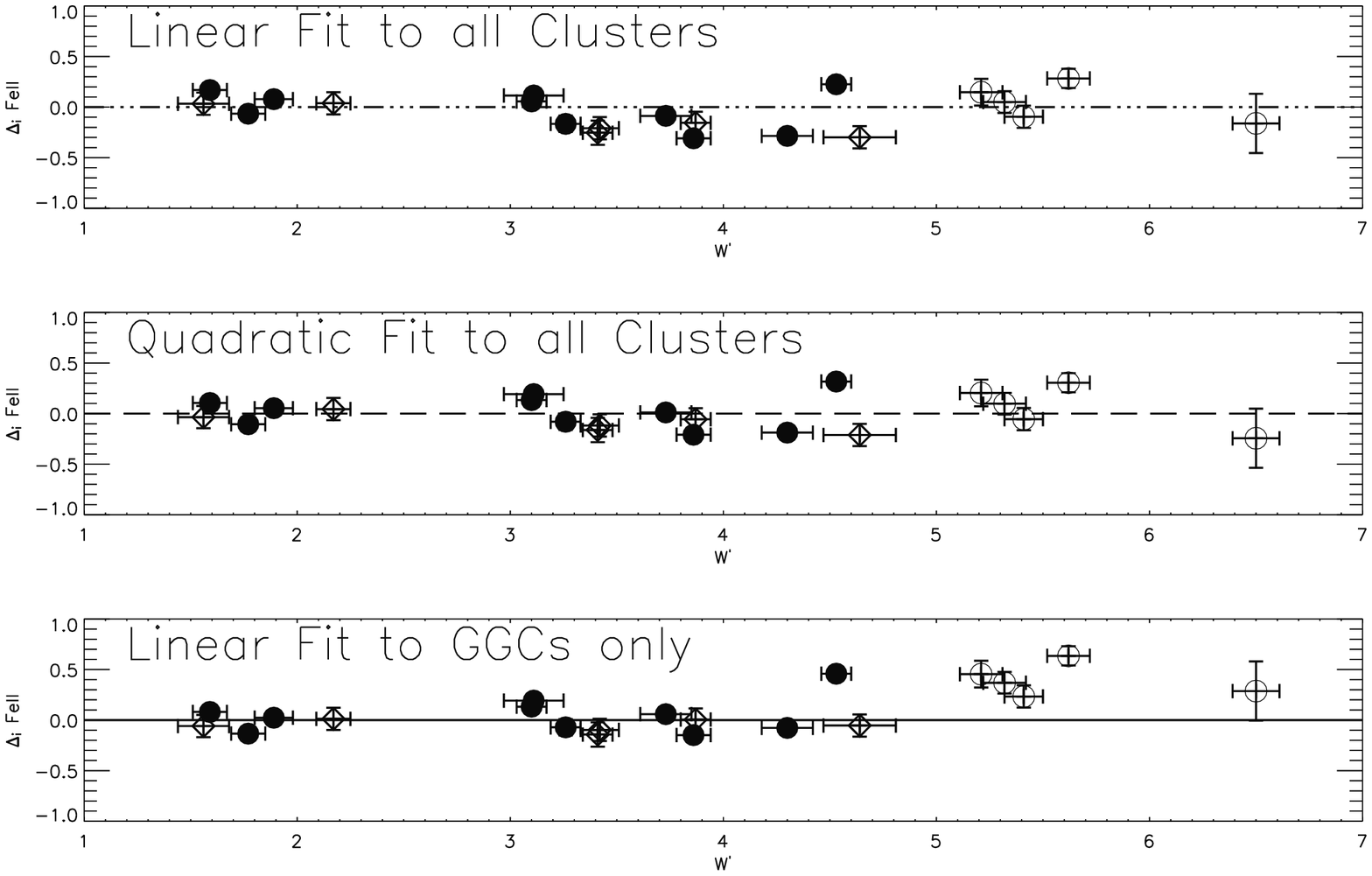}
            \caption[{[Fe/H]$_{II}$ versus \wp}]{{\small
Iron abundances from singly-ionized lines versus reduced
equivalent width and residuals for all of the clusters. Solid
circles indicate GGCs; open circles indicate GOCs; and the KI03
clusters scaled to this data set are indicated with open diamonds.
The dotted-dashed line indicates a linear fit to all stars, the
dashed line is a 2nd order polynomial, and the solid line
indicates a linear fit to the GGCs with [Fe/H] $<$ $-$0.8, see
text for details.
                   }  }
            \label{fig-wpfeII}
        \end{center}
    \end{figure*}

Bosler (2004) reinvestigated the widely-utilized relationship
between \wp and [Fe/H] for 21 Galactic cluster with a wide range
of ages (2 to 14~Gyrs) and metallicities (--2.5 $\le$ [Fe/H] $\le$
$+$0.2). \wp values, as mentioned in the previous section, were
measured for stars in the same 21 clusters to build an internally
self-consistent data set, and equivalent widths were transformed
to the scale of the R97a data because their Ca~II measurements are
so widely utilized. Figure \ref{fig-wpfeII} illustrates the
empirical relationship between \wp and [Fe/H] for the B04 clusters
along with the residuals for each of three different fits to the
data - a linear fit to only the globular clusters (analogous to
the R97b fit to the CG97 metallicity scale) and a linear and
quadratic fit to all clusters.

Note that the relationship found by R97b using the metallicity
scale of CG97, was based solely on Galactic globular cluster (GGC)
stars; the relationship they found, for comparison, was

\begin{equation}
\rm [Fe/H]_{CG97} = -2.66 (\pm 0.08) + 0.42 (\pm 0.02) \times
W^\prime.
\end{equation}

\noindent This fit is similar to the fit to our GGCs (see Table
\ref{fe_fits}). The relationship appears fairly smooth and linear,
but the addition of the younger, metal$-$rich Galactic open
clusters (GOCs) with [Fe/H] $>$ $-$0.8, indicates that [Fe/H] is
not truly a linear function of \wp. Since most work on the Ca~II
triplet calibration was done using GGCs, mostly with [Fe/H] $<$
$-$0.8, a liner fit is indicated by a solid line in Figure
\ref{fig-wpfeII}. If this calibration based on only the GGCs is
used to infer the abundances of the metal$-$rich open clusters the
derived metallicity would be lower by approximately 0.3~dex.

In fact, the relationship is linear up to [Fe/H] $\approx$
$-$0.08, but if all of the data are included, the relationship
becomes quadratic.  The linear fit to only the GGCs, the linear
fit to all clusters, and the quadratic fit to all of the clusters
are listed in Table \ref{fe_fits} along with the mean error of
unit weight, m.e.1\footnote{The m.e.1 $= \left(
\sum{\epsilon^2}/\sigma^2 \right)^{0.5}~(N - 1)^{-0.5}$, where
$\epsilon$ is the deviation of the [Fe/H] value for each cluster
from the predicted value based on the fit, $\sigma$ is the
uncertainty in \wp and [Fe/H] values, N is the total number of
clusters. A value of one indicates that the scatter is consistent
with the observational errors, and a higher value implies that the
errors have been underestimated or that the relationship is
nonlinear}, rms and the Chi-squared per degree of freedom,
${\chi_\nu}^2$, values for the regressions. Notice that linear fit
to all of the clusters has the largest m.e.1 value implying that
the either the errors have been underestimated or the fit to the
\wp to [Fe/H] relationship is really non-linear and is most likely
effected by the varying [Ca/Fe] values for the different clusters
(ie. it is dependent upon star formation history).

In order to circumvent the fundamental dependence on [Ca/Fe]
ratios built into the current Ca~II calibrations, B04 derived an
empirical relationship between Ca~II and [Ca/H].  It is important
to note that the [Ca/H] values {\it were not derived from the
Ca~II lines}.  The [Ca/H] values were determined from atmospheric
abundance analysis of neutral calcium (Ca~I) lines measured in
high-dispersion spectra.  The abundances derived for the Ca~I
lines were also corrected for non-LTE effects in the analysis.  In
B04, non-LTE effects on the Ca~I, Fe~I and Fe~II lines were
closely examined.  These effects on derived abundances arise from
the fact that some high energy UV photons from deep within the
stellar atmosphere penetrate the upper levels of the atmosphere
and break the LTE assumption used in the modelling the line
formation. The changes in temperature between the model and the
'true' atmosphere effects the abundances derived by the neutral
lines more strongly than the abundances derived by ionized lines.
As a result, abundances derived from Fe~II lines best represent
the 'true' abundances, while those from Fe~I and Ca~I lines will
be effected by the non-LTE atmosphere of the star. However, B04
found that the ratio of Ca~I to Fe~I ([Ca/Fe]$\rm_I$) is not
strongly effected by variations in stellar temperatures as both
species are being effect similarly. Therefore, the 'true' [Ca/H]
value was defined as

\begin{equation} \rm [Ca/H] = [Ca/Fe]_I + [Fe/H]_{II}, \end{equation}

\noindent where [Fe/H]$\rm_{II}$ is the iron abundance derived
from Fe~II lines.  This 'true' value of [Ca/H] is not affected by
the non-LTE effects described above.

The calibration is based on stars in 15 Galactic globular and open
clusters with ages from $\approx$ 2 to 14~Gyr and -2.2 $<$ [Ca/H]
$<$ +0.2 making it an extremely robust calibration of the Ca~II
triplet to abundances. The regression between \wp and [Ca/H] is

\begin{equation}
\rm [Ca/H] = -2.778 (\pm 0.061) + 0.470 (\pm 0.016)~dex/\AA\
\times W^\prime.
\end{equation}

This relationship is independent of the ages of the clusters for
clusters older than a few Gyrs. It can be applied to Galactic and
extragalactic stars without concern for the effects of variations
in [Ca/Fe] ratios in the stars.  Therefore, it is independent of
the star formation history of the system. Additionally, the new
calibration includes stars from from both GGCs and GOCs to insure
that it is valid over a wide range of ages and metallicities.

\begin{figure*}
            \includegraphics[width=\figwidth]{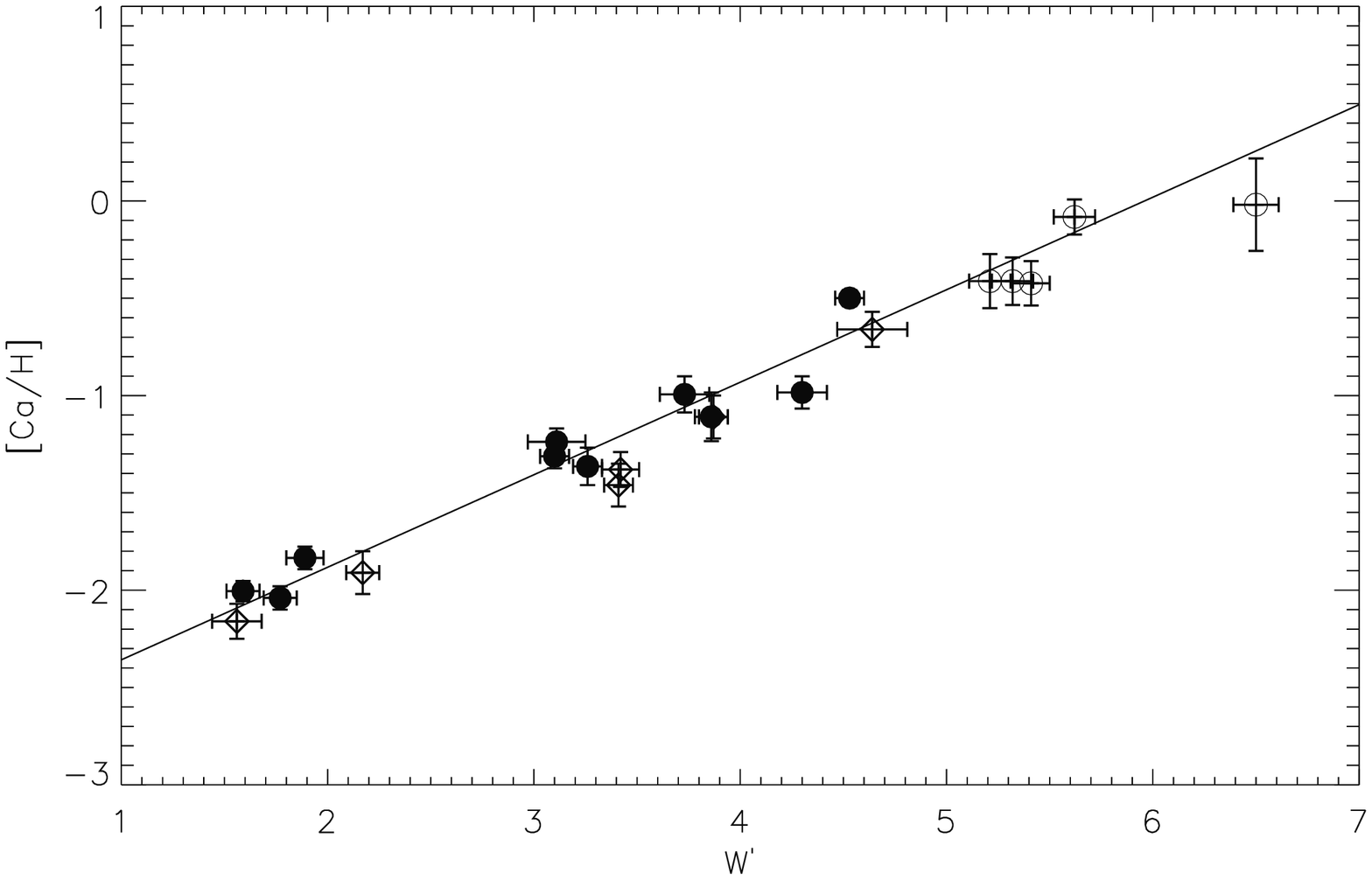}
            \includegraphics[width=\figwidth]{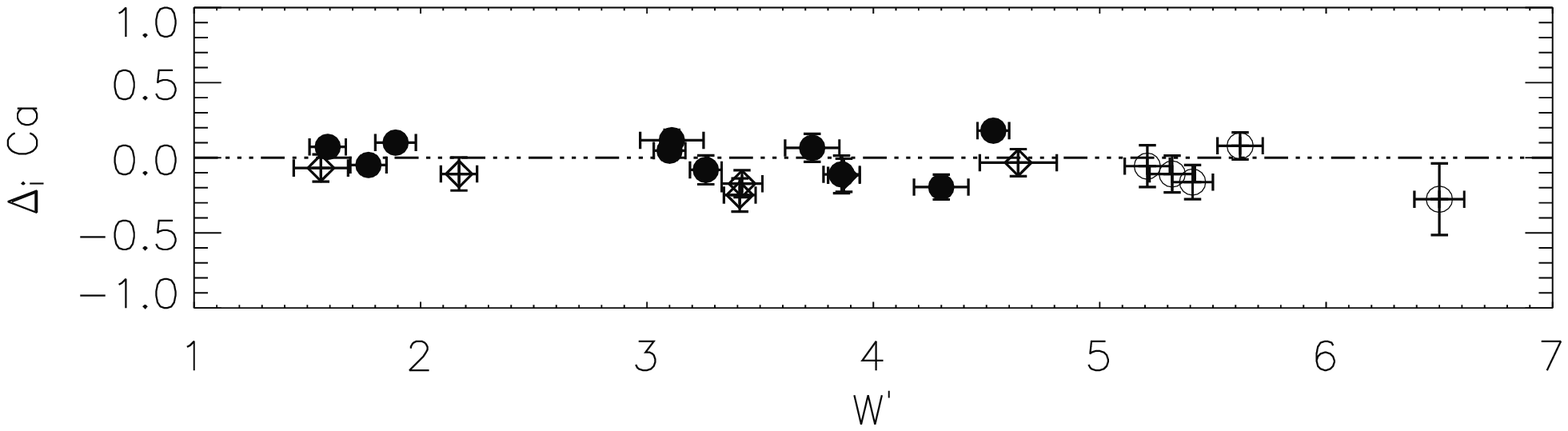}
            \caption{[Ca/H] versus \wp and residuals for all observed clusters.
            Solid circles indicate GGCs; open circles indicate GOCs; and the
KI03 clusters scaled to this data set are indicated with open
diamonds}
    \label{fig-cacalib}
    \end{figure*}

Figure \ref{fig-cacalib} shows the linear fit to the calibration
with an root-mean-square (rms) scatter of 0.13~dex, an mean error
of unit weight (m.e.1) $=$ 1.06.  The residual to the fit is also
illustrated.  The Chi-squared per degree of freedom,
${\chi_\nu}^2$ = 1.87. We also applied a quadratic fit to the data
and found that the coefficient on $\rm {W^\prime}^2$ is
statistically insignificant compared to the uncertainties in the
fit, which are over twice as large.  The values of Ca~II and
[Ca/H] for clusters used in the calibration are shown in Table
\ref{wp_abund}.  Note that the [Ca/H] values listed were derived
from atmospheric abundances analysis of high dispersion spectra,
while the value of \wp were derived from low-dispersion spectra of
stars in the same clusters.

\subsection{Which is Preferable - A Ca~II Calibration to [Ca/H] or
[Fe/H]?}

As the star formation histories and the effect of galactic winds
in the Leo~I and Leo~II are not yet well-defined, we do not yet
know what the element ratio, [Ca/Fe], is as a function of
metallicity.  Thus it would be inappropriate to assume that they
follow the same [Ca/Fe] verses [Fe/H] relationship as that found
for the Galactic Globular clusters used by R97 to calibrate the
Ca~II equivalent widths to [Fe/H].  The new calibration to [Ca/H]
has a very small RMS that can be explained mostly by observational
error rather than systematic variations that depend on [Ca/Fe] and
is also independent of stellar ages (for ages~$\ge$~3~Gyrs) and
star formation history. Therefore, the new calcium calibration
will be used to derive abundances in this work; however, [Fe/H]
values on the scale of CG97 will also be listed for reference to
those still compelled to use the existing calibration.

As an aside, there is really no compelling reason that stellar
metallicities should be specified in terms of [Fe/H].  The iron
abundance became the observational spectroscopist's way of
characterizing the overall metallicity of a star in a
one-dimensional way, because iron and iron-peak elements are the
most plentiful lines and are easiest to measure in optical and
near-IR spectra. In fact, stellar opacities are controlled by the
entire range of elements; the iron and iron-peak elements whose
outer electrons are most easily ionized, but, more importantly,
the CNO and alpha-elements, which are most abundant by number
highest and which provide the majority of the transition
electrons. Additionally, the abundance ratios [CNO/H] in high
metallicity stars is an important parameter because it also
controls the amount of energy generated from the CNO cycle.

\section{Results}

\subsection{Metallicity Distribution of Leo~I}

The observed range in reduced equivalent widths for Leo~I dSph stars
is $\rm 0.83 \le W^\prime \le 4.63 $~\AA. Tables~\ref{table-leo1}
and \ref{table-leo1_spec} list the parameters of each star (V, V--B,
Ra, Dec, SNR and V$_{Helio}$) and the spectroscopic values (\wp,
[Fe/H]$_{CG97}$, and [Ca/H]), respectively. Note that the \wp values
have been scaled to the work of R97a, which has been used as a
standard in the literature.

\begin{figure*}
           \includegraphics[width=\figwidth]{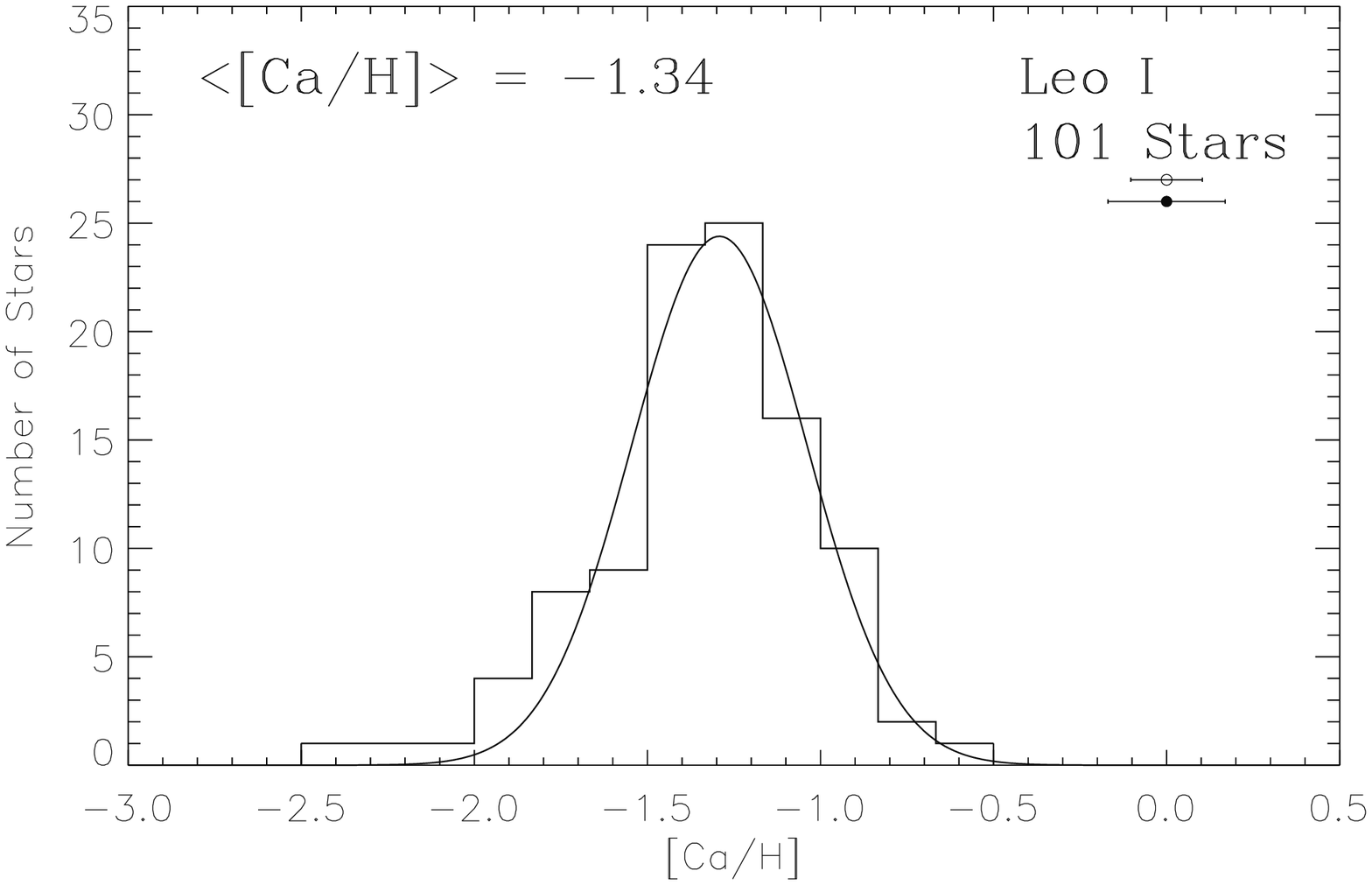}
           \caption{The metallicity distribution function in the Leo~I dSph.
Typical 1-$\sigma$ errors shown in the upper right;  random and
total errors are shown as open and filled circles, respectively.
The Gaussian fit to the distribution is shown as a solid line.}
            \label{fig-leo1_caspread}
            \end{figure*}

As shown in Figure~\ref{fig-leo1_caspread}, the inferred
metallicities range from $\rm -2.39 \le [Ca/H] \le -0.60$ for 102
stars. The average metallicity is [Ca/H] = $-1.34 \pm 0.02$, and the
observed spread is 1.20~dex. The median random [Ca/H] error is
0.10~dex, and the median total error is 0.17~dex per star. Placing
our Ca~II values on the [Fe/H] scale of CG97, Leo~I has a mean
[Fe/H]$\rm_{CG97} =-1.34$.  Calcium abundances have not been studied
extensively in Leo~I. However, Shetrone \etal (2003) found [Ca/Fe] =
$+$0.15 and $+$0.02~dex ($\sigma \approx$ 0.15~dex) and [Fe/H] =
$-$1.52 and $-$1.06~dex, respectively, ($\sigma \approx$ 0.1~dex)
for two stars in Leo~I based on abundance analysis of
high-dispersion spectra. This implies [Ca/H] = $-$1.20, which is
slightly more metal-rich than the average we found for 102 stars,
but the intrinsic spread we see and the small number of stars used
by Shetrone \etal could account for the difference.

Suntzeff \etal (1986) found an average metallicity of [Fe/H] =
$-$1.5 $\pm$ 0.25 from the strengths of the Ca H and K lines in low
dispersion spectra of two red giants, and this corresponds to [Ca/H]
= $-$1.41 assuming an average [Ca/Fe] from Shetrone \etal (2003)'s
two stars.  This value is more metal-poor than the Shetrone \etal
value, and it is in agreement with our observed values given the
small number of stars used.  More recently, Koch \etal (2006b)
examined a very similar sample of Leo~I stars and derived [Fe/H]
from the Ca~II triplet for 58 red giant stars and found
[Fe/H]$_{CG97}$ = $-$1.31$\pm$0.02 and a metallicity spread of
1~dex, which agrees very well with our observations.  The lack of
overlap between the two data sets does not allow for comparison of
individual stars, but the trend in the Leo~I abundances is constant
between the two data set.

Leo~I has a roughly Gaussian-shaped metallicity distribution with an
asymmetric over-abundance at the metal-rich end and a slightly
extended metal-poor tail. This distribution is very similar to that
of Koch \etal (2006b) with a slightly higer resolution due to the
larger number of stars used in this sample.  We fit the metallicity
distribution function with a Gaussian distribution of the form:

\begin{equation}
\rm N(m) \propto exp \left( - \frac{(m-m_0)^2}{2\sigma_m^2}
\right) ,
\end{equation}

\noindent where $m =$ [Ca/H], in order to determine the centroid
of the distribution, $m_0$, and the observed dispersion,
$\sigma_m$. For Leo I, a non-linear least-squares fit gives $m_0 =
-1.35 \pm 0.02$, and $\sigma_m = 0.24 \pm 0.03$. The fit is shown
in Figure~\ref{fig-leo1_caspread}. Given that the median random
measurement error is $\sigma_{\rm rand} = 0.10 \pm 0.03$~dex, the
intrinsic metallicity dispersion in Leo~I is equal to

\begin{equation}
\sigma_{m,i}  = (\sigma_m^2 - \sigma_{\rm rand}^2 )^{\frac{1}{2}}
=
  0.26 \pm 0.03 ~{\rm dex}
\end{equation}

\noindent The implied intrinsic spread in metallicity in Leo~I is
approximately 2 FWHM $= 4.71~\sigma_{m,i} = 1.22 \pm $~dex.

\begin{figure*}
            \includegraphics[width=\figwidth]{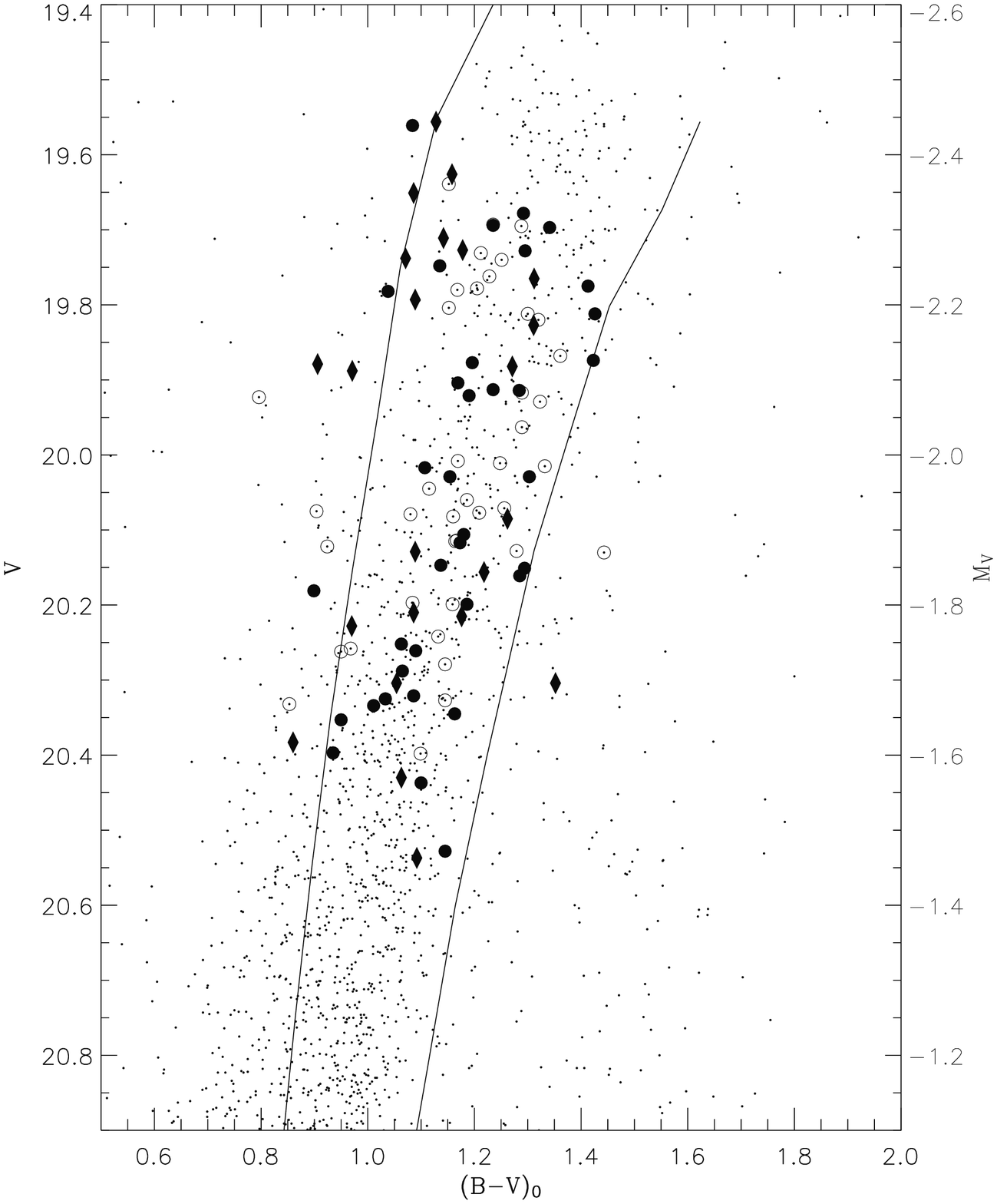}
            \caption{The CMD of Leo~I showing the metallicities of the stars.
             V, B--V rather than I, V-I is shown for ease of comparison
             with the Leo~II CMD.  Stars are coded based on their derived metallicity: filled diamonds
for stars with $\rm {[Ca/H]<-1.5}$; open circles for stars with
$\rm{-1.5\leq[Ca/H]<-1.2}$; filled circles for stars with
$\rm{[Ca/H] \ge -1.2}$.  The fiducials for GGCs M5 ([Ca/H] =
--0.96) and M68 ([Ca/H] = --1.78) are over-plotted for reference.}
            \label{fig-leo1_RGB}
            \end{figure*}

Figure \ref{fig-leo1_RGB} shows the position of stars in the CMD
as a function of their metallicity.  The fiducials for the GGCs M5
= NGC~5904 ([Ca/H] = $-$0.96, age = 14~Gyr) and M68 ([Ca/H] =
$-$1.78, age = 14~Gyr) are plotted for reference in the CMD. The
most metal-poor stars tend to lie along the blue half of the RGB,
but stars with [Ca/H] $\simgt -1.5$ occupy a wide swath in color.
At any given color, there are stars with a mix of metallicities,
which probably reflects the fact that the red giants in Leo~I have
a large range in ages (many Gyrs).

A large spread in metallicity has not been assumed in most studies
of the Leo~I CMD (Lee, et al.~1993, Gallart, et al.~1999b, Siegel
\& Majewski~2000), which either assumed a constant metallicity or
a much narrower range. Including a 1.2~dex metallicity spread will
dramatically affect the inferred star-formation history.

Dolphin~(2000) did include metallicity evolution as a variable in
his CMD modelling of the Leo~I and Leo~II CMDs. He photometered
archival WFPC2 images and obtained a CMD that spanned the tip of
the RGB to the oldest main-sequence turnoff. (The photometric
errors at the level of the oldest MSTO are much larger in the
Leo~I data, and hence the SFH for ages $\simgt 9$~Gyr in the Leo~I
dSph was not as well determined as for the Leo~II dSph.) Dolphin
modelled the Hess diagram-the density of stars in the CMD-using
Padova isochrones (Girardi, et al.~2000), which he interpolated on
a finer grid of metallicity (0.15~dex binning). He has kindly
provided the simulated data from his best fit models.

\begin{figure*}
            \includegraphics[width=\figwidth]{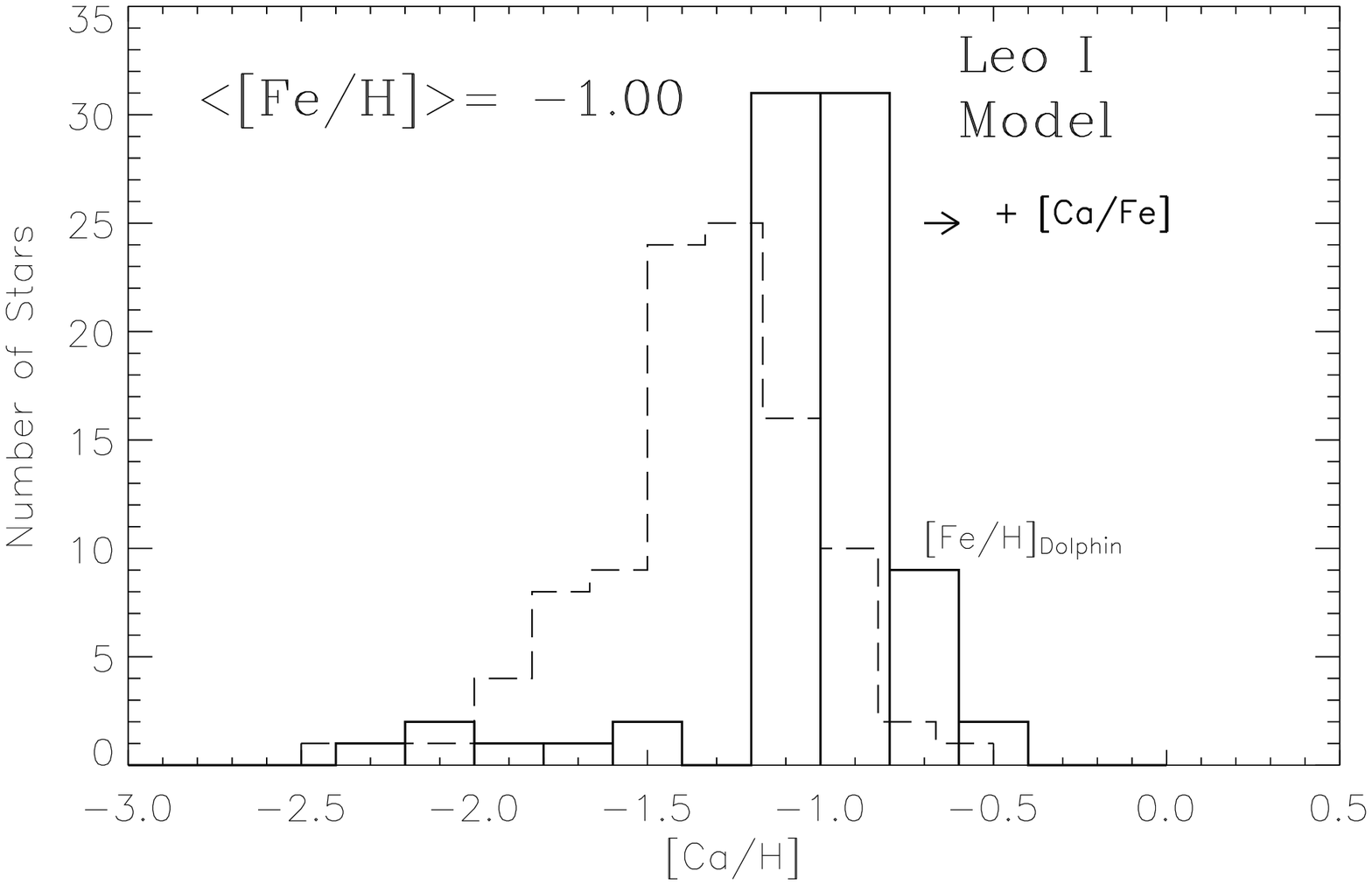}
            \caption{The predicted metallicity distribution from Dolphin's CMD
            modelling of the Leo~I dSph (solid lines) along with the observed distribution
            (dashed lines).  Note that Dolphin derived [Fe/H] and a significant offset
            exists between the two distributions.  If [Ca/Fe] from Shetrone \etal (2003)
            is added to Dolphin's [Fe/H] values yield [Ca/H] (indicated with an arrow of length 0.1~dex),
            the theoretical histogram shifts even further from the observed distribution
            (see text for discussion).}
            \label{fig-dolphindata}
            \end{figure*}

We have applied to Dolphin's model CMD the same color and magnitude
selection criteria used to select the Leo~I spectroscopic sample
(note that the color selection did not eliminate any RGB stars),
computed the metallicity distribution, and convolved it with our
median random [Ca/H] errors.  The resulting metallicity distribution
function is shown in Figure~\ref{fig-dolphindata}.  The shapes of
the model and observed distributions are similar, with both having
slight excesses at the metal-rich end and a metal-poor tail,
however, there is a deficiency in metal-poor stars in the model
distribution compared to the observations. More recent photometry
from Held \etal (2000) indicates a significant metal-poor population
based on the existence of an extended blue horizontal branch that
had not been previously observed.  The exclusion of this metal-poor
population in the synthetic CMD used to predict the dispersion may
account for the missing metal-poor stars. Interestingly, the peak of
the model is significantly more metal-rich than the observed
distribution. A Gaussian fit to the model gives $m_0 = -1.0$ and
$\sigma_m=0.17$~dex, although a Gaussian is obviously a poor
assumption for this model distribution.

Dolphin did examine [Fe/H] values rather than [Ca/H], but the peak
of the distribution does not account for potential [Ca/Fe]
differences that may exist in the stars. What could the be cause of
the discrepancy? In the CMD modelling, the best fitting values of
the metallicity and star-formation rate as a function of time are
determined by matching the density of stars in finely-binned
magnitude and color intervals in the observed and modelled CMD. In
the process, the portions of the CMD with the largest numbers of
stars are given the greatest weight. These are the main-sequence
turnoff and unevolved main-sequence regions.  The width in color of
the main-sequence turnoff and upper main sequence is governed by the
spread of ages, metallicities, and the assumed fraction and
characteristics of unresolved binary systems. A high fraction of
binaries causes a spread in the color of the unevolved main
sequence, which could be interpreted as an intrinsic spread in
metallicity. Thus the assumptions made about the unresolved binaries
could affect the inferred distribution of metallicities. In
addition, the models themselves could be source of problems because
the color-temperature relationships used to transform from the
theoretical to observed plane are still controversial.

More recently, Koch \etal (2006c) examined simple chemical evolution
models to the observed Leo~I distribution.  They used a modified
simple closed-box model and a model using Prompt Initial Enrichment
scaled to the number of stars they observed.  Their model
distributions yielded similar average abundances for Leo~I, but the
closed-box model severally over estimated the metallicity spread
while underestimating the number of stars near the center of the
distribution.  The Prompt Initial Enrichment model did a better job
at predicting the height of the distribution peak, but it
underestimated the metallicity spread in the galaxy.  See in Koch
\etal (2006c) (Figure 12) for a more detailed discussion.

\subsection{Metallicity Distribution of Leo~II}

\begin{figure*}
            \includegraphics[width=\figwidth]{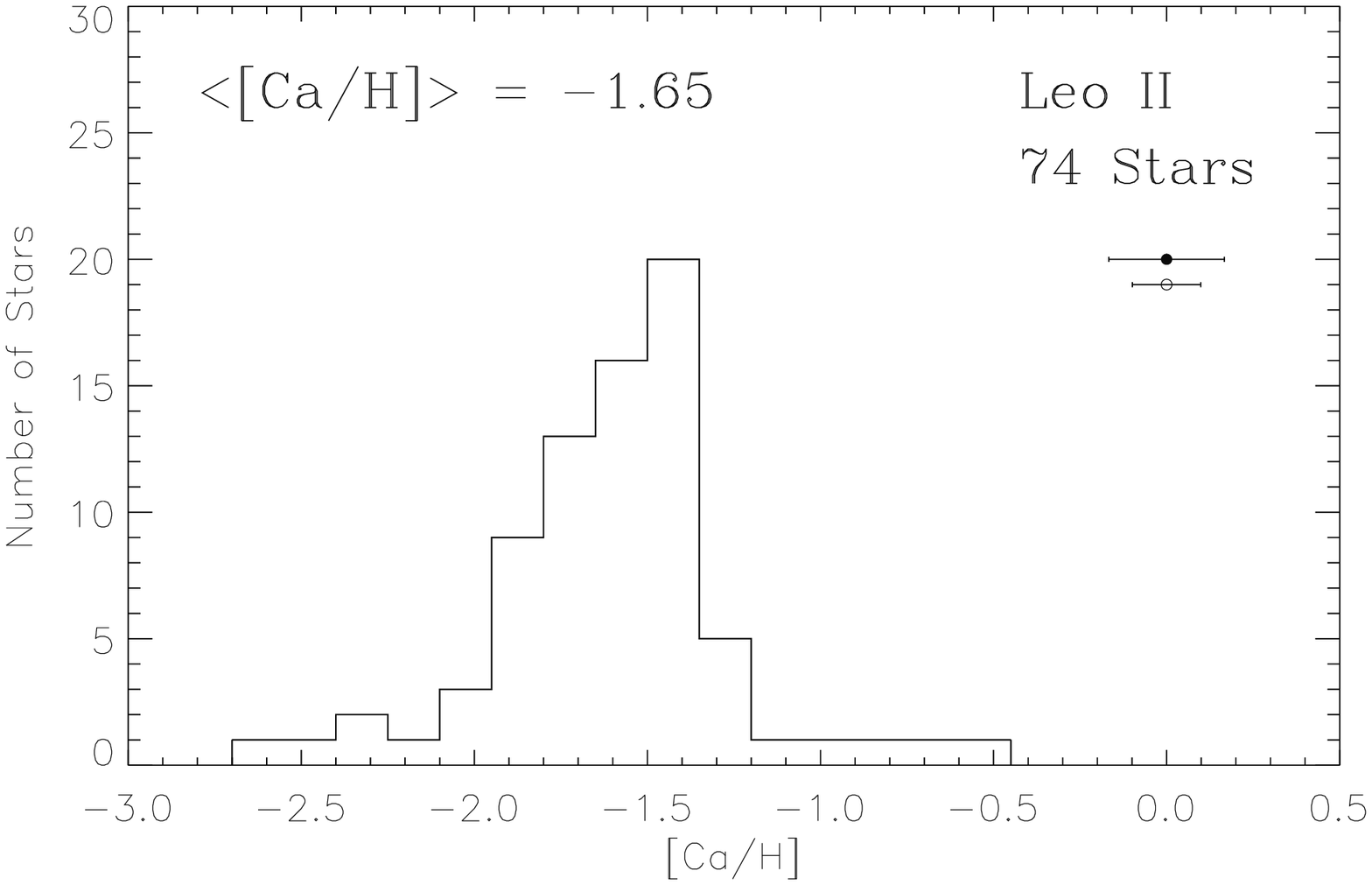}
            \caption{The observed metallicity distribution function in the Leo~II
dSph. Typical 1-$\sigma$ errors are shown in the upper right;
median random and total errors are shown as open and filled
circles, respectively. }
            \label{fig-leo2_caspread}
            \end{figure*}

The observed range in the reduced equivalent widths for Leo~II stars
is $\rm{0.36 \le W^\prime \le 4.65}$~\AA. Tables\ref{table-leo2} and
\ref{table-leo2_spec} list the parameters of each star (V, V--B, Ra,
Dec, SNR and V$_{Helio}$) and the spectroscopic values (\wp,
[Fe/H]$_{CG97}$, and [Ca/H]), respectively.  Note that the \wp
values have been scaled to the work of R97a.
Figure~\ref{fig-leo2_caspread} shows the range in metallicities is
$\rm -2.61 \le [Ca/H] \le -0.59$, and the observed spread is
1.29~dex. The average metallicity in the Leo~II dSph is [Ca/H] =
$-1.65 \pm 0.02$, much more metal-poor than the median in the Leo~I
dSph. Placing our Ca~II values on the [Fe/H] scale of CG97, Leo~II
has a mean [Fe/H]$\rm_{CG97} =-1.59$. The median random error is
0.10~dex and the median total error is 0.17~dex per star, and the
intrinsic metallicity dispersion is 1.01~dex. Leo~II also has an
asymmetric distribution with a steep cut off on the metal-rich end,
excesses at the metal-poor end, and an indication of metal-rich and
metal-poor tails.

Suntzeff, et al.~(1986) found an average $\rm [Fe/H] = -1.9 \pm 0.2$
based on the Ca H and K strengths of 3 red giants.   With an unknown
typical [Ca/Fe] value for the Leo~II stars, it is impossible to
directly compare the two results.  However, assuming an average
[Fe/H] $\approx$ --1.90 from Suntzeff and our [Ca/H] would imply
[Ca/Fe] $\approx$ $+$0.3, which is similar to the values for older
stars in Milky Way globular clusters and field stars. Mighell \&
Rich~(1996) inferred an average $\rm [Fe/H] = -1.60 \pm 0.25$ using
the shape of the RGB in the V, I-band CMD. From interpolating
globular cluster fiducial RGBs, they derived an intrinsic spread in
[Fe/H] of 0.9~dex, which is similar to the spread in [Ca/H] of
1.0~dex derived here. This agreement is understandable because the
color and shape of the RGB are much more sensitive to metallicity
than age for ages $\simgt 7$~Gyr, the age of the youngest stars in
the Leo~II dSph.  Koch \etal (2006b) observed 52 Leo~II stars and
found [Fe/H]$_{CG97}$ = $-1.74 \pm 0.02$ and a spread of 1.3~dex.
Though the average metallicity is more metal-poor than out
observations, the distribution of the stars in both data sets are
very similar (see Figure~\ref{fig-leo2_caspread} in this paper and
Figure~7 in Koch \etal 2006b for comparison).

\begin{figure*}
            \includegraphics[width=\figwidth]{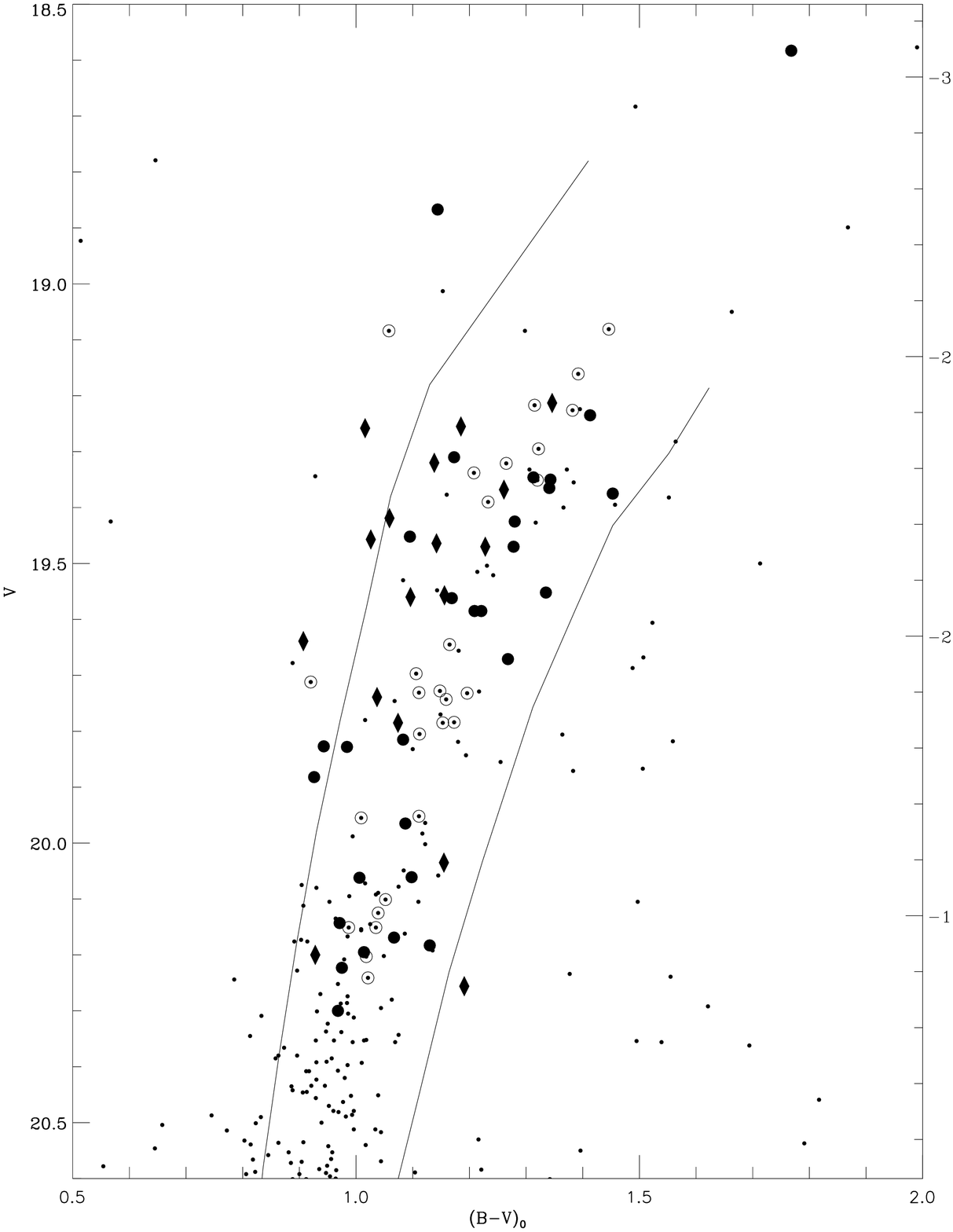}
            \caption{The CMD of Leo~II showing the metallicities of the stars. Stars
are coded based on their derived metallicity: filled diamonds for
stars with $\rm{[Ca/H]<-1.8}$; open circles for stars with
$\rm{-1.8\le[Ca/H]
> -1.5}$; filled circles for stars with $\rm{[Ca/H] \geq -1.5}$
The fiducials for GGCs M5 ([Ca/H] = --0.96) and M68 ([Ca/H] =
--1.78) are over-plotted for reference. }
            \label{fig-leo2_RGB}
            \end{figure*}

Figure \ref{fig-leo2_RGB} illustrates the range in [Ca/H] across the
RGB. Again, the large dispersion in metallicity at a given color
indicates the stars span a range of ages.

\begin{figure*}
            \includegraphics[width=\figwidth]{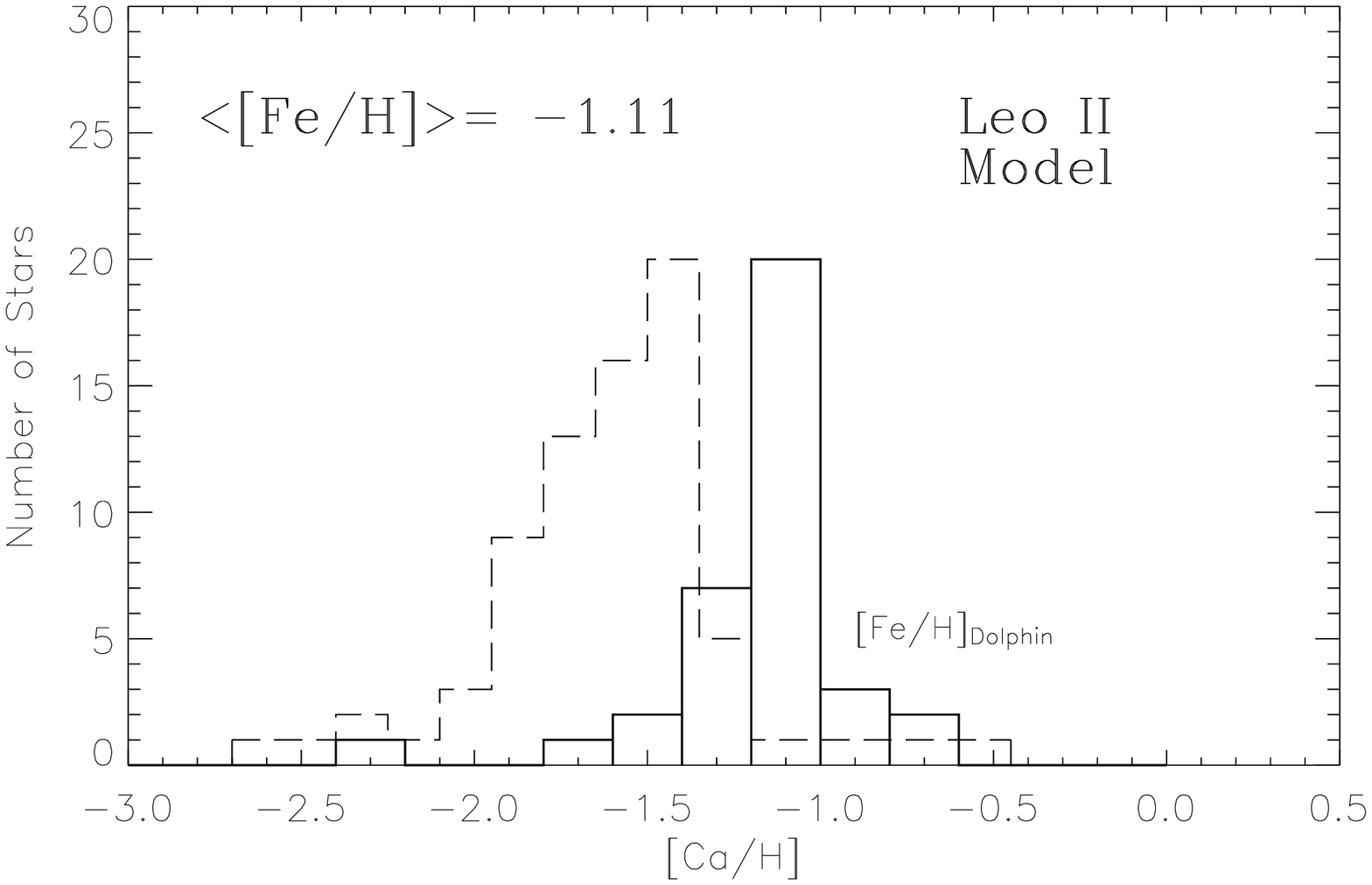}
            \caption{The predicted metallicity distribution from Dolphin's CMD
            modelling of the Leo~II dSph (solid lines) along with the observed distribution
            (dashed lines).  Note that Dolphin derived [Fe/H] rather than [Ca/H] values, and
            a significant offset exists between the central values of the two distributions
            (see text for discussion).}
            \label{fig-dolphindata_leo2}
            \end{figure*}

We also compared the observed metallicity distribution of the
Leo~II dSph with predictions of CMD modelling from Dolphin (2002).
Since both the model and observations have a very non-Gaussian
distribution, we did not attempt to compute Gaussian fits.
Figures~\ref{fig-leo2_caspread} and \ref{fig-dolphindata_leo2}
show that the shape of the model and observed distributions are
similar with excesses at the metal-poor end and a steeper drop at
the metal-rich end.  Leo~II is the smaller of the two dSphs,
having $\rm M_V = -9.6$ compared to $\rm M_V = -11.9$ for Leo~I.
The decreased number of stars at high metallicities in Leo~II may
be due to its greater inability to retain gas after star formation
had progressed for a while. Analyzing the CMD from the same set of
WFPC2 images, Mighell \& Rich (1996) and Dolphin (2002) both
inferred that Leo~II stopped forming stars $\approx$ 7~Gyr ago,
compared to the Leo~I dSph which only stopped forming stars $\sim
2$~Gyr ago.

Though the shapes of the distributions are similar, the model has
a median which is more metal-rich than the median of the observed
distribution.  Assuming the [Ca/H] from this work and [Fe/H]
values from Dolphin are taken to be typical values of the Leo~II
stars, it would imply a [Ca/Fe] $\approx$ --0.49.  The large
discrepancy between the abundances derived from CMD modelling and
the observations also exists for Leo~II stars, and it is even
larger than those of the Leo~I stars and much lower than the
predicted abundance ratios from Type Ia supernovae. Again, if we
compare our [Fe/H]$\rm_{CG97}$ values with those of Dolphin, the
discrepancy still exists, and is approximately the same size.
Though there are no available data on abundance ratio of Leo~II,
the extremely low [Ca/Fe] value from comparing the model results
is dramatically different than those derived by comparing [Ca/H]
to spectroscopic [Fe/H] abundances (Shetrone \etal 2003), which
yields [Ca/Fe] more typical of Galactic and Local Group stars.

Koch \etal (2006b) compared a simple closed-box chemical evolution
model (valid only for long-lived stars) with their Leo~II
distribution and found agreement with central [Fe/H] values, but the
model severely underestimated the number of stars in the central
metallicity range and over estimated the metallicity spread by over
estimating the number of metal-poor stars in the galaxy.
Additionally, Koch \etal (2006b) compared the models from Lanfranchi
and Matteucci (2004) for other Local Group dSphs to the observed
Leo~II distribution.  The Carina, Draco, Sagittarius, Sextans,
Sculptor and Ursa Minor dSphs were compared, and they found that
Leo~II and Carina were the two most similar dSphs in the set using a
very simplified comparison.  Not surprisingly, they found that the
Leo~II SFH is as unique and distinct from the other dSphs in the
Local Group, reflecting the fact that one global explanation for the
formation of these seemingly simple galaxies does not reflect their
observed complexities.

\section{Summary}

We have determined calcium abundances for 102 RGB stars in the
Leo~I dSph galaxy and 74 RGB stars in the Leo~II dSph galaxy using
the equivalent widths of the near infrared Ca~II triplet lines.
The metallicity scale used to convert the Ca~II measurements is
based on our calibration of Ca~II to [Ca/H] from Galactic cluster
stars with ages from $\approx$ 2 to 14~Gyr and $-2.2$ \ltsima
[Ca/H] \ltsima $+0.2$. Unlike the calibration of Ca~II to [Fe/H]
developed by R97b, the Ca~II to [Ca/H] is independent of
star-to-stars variations in [Ca/Fe] with [Fe/H]. Defying the
tradition of using ``metallicity" interchangeably with iron
abundances, [Ca/H] was taken as the relevant metallicity in this
investigation.

The Leo~I dSph has a average metallicity of [Ca/H] = $-1.34 \pm
0.02~ (\sigma = 0.21)$ and an intrinsic spread in metallicity of
1.2~dex. The average metallicity of the Leo~II dSph is more
metal-poor with [Ca/H] = $-1.65 \pm 0.02~ (\sigma = 0.17)$ and has
a 1.0~dex spread in metallicity.

The metallicity distribution function is different for the two
dSphs. Leo~I has an approximately Gaussian distribution with a
slight excess at the metal-rich end, while Leo II has a steeper
cutoff at the metal-rich end at [Ca/H] $\approx -1.4$. The
differences in the shape of the distributions may be caused by
greater mass loss in the less massive Leo~II dSph, whose star
formation ended approximately 7~Gyr ago, compared to the more
massive Leo~I dSph, whose star formation ended only $\sim 1$ to
2~Gyr ago. Significant differences, on the order of $-$0.5~dex,
exist between the observed metallicity distribution functions and
those derived by sophisticated modelling of Hess diagrams in CMDs
obtained with HST. However, the assumption of a Galactic range of
[Ca/Fe] values show our results to be consistent with the [Fe/H]
values derived in other spectroscopic and photometric studies.

\section{Acknowledgments}

We gratefully acknowledge financial support from the NSF through
grant AST-0070895 to T. Smecker-Hane. T. Bosler would like to
thank the ARCS Foundation for a graduate student fellowship and
Sigma Xi for a Grant in Aid of Research (2001) as well as ANdy
McWilliam, Virginia Trimble, Jay Gallagher and Linda Sparke for
their inspiration and interest in this publication. We thank the
staff at Keck Observatory, in particular, Paola Amico and Greg
Wirth, for their excellent support, and we thank Bev Oke, Judy
Cohen and their collaborators for an extremely useful
spectrograph. In addition, we extend special thanks to the people
of Hawai`ian ancestry on whose sacred mountain, Mauna Kea, we were
privileged to be guests. Our research was expedited by the wide
range of data made available at Canadian Astronomy Data Center
(CADC), which is part of the Herzberg Institute of Astrophysics, a
facility of the National Research Council of Canada. P. Stetson is
extremely grateful to Howard~E.~Bond, Enrico~V.~Held,
Giampaolo~Piotto, Edward~W.~Olszewski, Ivo~Saviane,
Robert~A.~Schommer, and Nicholas~B.~Suntzeff, as well as to the
CADC and the Isaac Newton Group Archives, for providing much of
the CCD imagery used in this paper.

{}
\clearpage

%INSERT TABLES - \clearpage between tables
\begin{table*}
\caption{Galaxy Properties} \vspace{0.1in}
    \begin{tabular}{llll}
    \hline \hline
& Leo~I & Leo~II & \\
\hline
 Right Ascension (J2000) & $\alpha =$ 10:08.5    & $\alpha =$11:13.5  & 1 \\
 Declination (J2000)     & $\delta =$ $+12$:18.5 & $\delta =$ $+22$:09.2 & 1 \\
 Total Absolute Magnitude, $\rm M_V$ (mag) & $-11.9$ & $-9.6$ & 1 \\
 Core and Tidal Radii (arcmin) &  3.3, 12.6 & 2.9, 8.7 & 1 \\
 Distance (kpc)          & 273 & 204 & 2, 3  \\
 Distance Modulus, $\rm (m - M)_0$ (mag) & 22.18 & 21.55 & 2, 3 \\
 Horizontal Branch Magnitude, $\rm V_{HB}$ (mag) & 22.70 & 22.17 & 4, 5 \\
 Foreground Reddening, E(B$-$V) (mag) & 0.040 & 0.027 & 6, 3 \\
    \hline \hline
    & & &\\
\multicolumn{4}{l}{\parbox{5.5in}{\renewcommand{\baselinestretch}{1.}
\normalsize{(1) Mateo 1998; (2) Lee, et al.~1993; (3) Mighell \&
Rich 1998;
 (4) Held, et al.~2001; (5) Siegel \& Majewski 2000; (6) Schlegel,
 Finkbeiner \& Davis 1998.}}}\\
\end{tabular}
\label{table-dsphs}
\end{table*}

\clearpage

%{\small \begin{center}
\begin{table*}
\caption[]{Galactic Globular Cluster Stars}
 \vspace{0.1in}
    \begin{tabular}{lcccccl}
\hline \hline Cluster-Star ID & \vvhb & \sumw & \sigw & $\rm
\sum{W_{Rut97a}}$ & $\rm \sigma_{W_{\sum{Rut97a}}}$ & Ref. \\
 & (mag) &(\AA) & (\AA) &(\AA) & (\AA) & \\
\hline
NGC 1904 -  52   & $-$2.09 & 3.84 & 0.12 & 4.15  & 0.15 & 1, 2\\
NGC 1904 - 131   & $-$2.54 & 4.27 & 0.17 & 4.71  & 0.14 & 1, 2\\
NGC 1904 - 223   & $-$2.96 & 4.25 & 0.15 & 4.80  & 0.16 & 1, 2\\
NGC 1904 - 160   & $-$3.13 & 4.13 & 0.15 & \nodata & \nodata  & 1, 2\\
NGC 1904 - 181   & $-$2.20 & 3.72 & 0.11 & \nodata & \nodata  & 1, 2\\
NGC 1904 -  89   & $-$1.44 & 3.61 & 0.10 & \nodata & \nodata  & 1, 2\\
NGC 1904 - 241   & $-$2.54 & 4.36 & 0.16 & \nodata & \nodata  & 1, 2\\
NGC 1904 - 294   & $-$1.93 & 3.97 & 0.14 & \nodata & \nodata  & 1, 2\\
 & & & & & & \\
NGC 4590 - AL81    & $-$1.89 & 2.90 & 0.16 & 2.64 & 0.07 & 3, 5\\
NGC 4590 - HI239   & $-$1.42 & 2.31 & 0.07 & 2.62 & 0.07 & 4, 5\\
NGC 4590 - HI119   & $-$2.04 & 2.44 & 0.05 & 2.76 & 0.06 & 4, 5\\
NGC 4590 - HQ$^{a}$& $-$2.11 & 2.56 & 0.07 & 3.00 & 0.04 & 4, 5\\
NGC 4590 - HII47   & $-$0.55 & 2.18 & 0.13 & 2.36 & 0.10 & 4, 5\\
NGC 4590 - HII28   & $-$1.97 & 2.90 & 0.05 & 2.99 & 0.06 & 4, 5\\
NGC 4590 - HI30    & $-$1.53 & 2.50 & 0.10 & 2.70 & 0.07 & 4, 5 \\
NGC 4590 - HI35    & $-$1.13 & 2.33 & 0.11 & 2.24 & 0.10 & 4, 5\\
NGC 4590 - 38      & $-$1.11 & 2.12 & 0.05 & \nodata & \nodata   & 5, 6\\
NGC 4590 - 31      & $-$1.32 & 2.23 & 0.06 & \nodata & \nodata   & 5, 6\\
NGC 4590 - 36      & $-$1.10 & 2.13 & 0.04 & \nodata & \nodata   & 5, 6\\
NGC 4590 - 22$^{a}$& $-$1.59 & 4.58 & 0.13 & \nodata & \nodata  & 5, 6\\
NGC 4590 - HI82    & $-$2.99 & 3.26 & 0.08 & \nodata & \nodata  & 4, 5\\
NGC 4590 - 23      & $-$1.48 & 2.14 & 0.05 & \nodata & \nodata  & 5, 6\\
NGC 4590 - A53     & $-$2.78 & 2.98 & 0.08 & \nodata & \nodata  & 3, 5\\
NGC 4590 - HI49    & $-$1.08 & 1.84 & 0.11 & \nodata & \nodata  & 4, 5 \\
 & & & & & & \\
NGC 5272 - 265   & $-$2.39 & 4.541 & 0.137 & 4.47  & 0.12 & 7, 8\\
NGC 5272 - 334   & $-$2.41 & 4.487 & 0.127 & 4.73 & 0.16 & 7, 8\\
NGC 5272 - 640   & $-$2.38 & 4.488 & 0.126 & 4.28  & 0.07 & 7, 8\\
NGC 5272 - 885   & $-$2.18 & 4.073 & 0.118 & 4.31 & 0.11 & 7, 8\\
NGC 5272 - 1217  & $-$1.65 & 4.115 & 0.091 & \nodata & \nodata   & 7, 8\\
NGC 5272 - 589   & $-$2.78 & 4.624 & 0.129 & \nodata & \nodata   & 7, 8\\
NGC 5272 - 250   & $-$1.54 & 3.965 & 0.111 & \nodata & \nodata   & 7, 8\\
NGC 5272 - 238   & $-$3.03 & 4.679 & 0.099 & \nodata & \nodata   & 7, 8\\
NGC 5272 - 345$^{a}$
                 & $-$2.19 & 4.513 & 0.124 & \nodata & \nodata & 7, 8 \\
 & & & & & & \\
NGC 6171 - SL   & $-$1.66 & 5.00 & 0.16 & 4.75 & 0.11 & 5, 6 \\
NGC 6171 - SR   & $-$1.04 & 4.25 & 0.14 & 4.27 & 0.14 & 5, 6\\
NGC 6171 - S62  & $-$1.73 & 4.81 & 0.14 & 5.13 & 0.07 & 5, 6\\
NGC 6171 - SF   & $-$2.31 & 5.08 & 0.13 & 5.45 & 0.09 & 5, 6\\
NGC 6171 - SS   & $-$0.91 & 4.72 & 0.16 & 4.88 & 0.15 & 5, 6\\
NGC 6171 - SU   & $-$0.92 & 4.04 & 0.10 & 4.44 & 0.22 & 5, 6\\
NGC 6171 - SH   & $-$1.86 & 5.01 & 0.14 & 5.16 & 0.09 & 5, 6\\
NGC 6171 - S203 & $-$0.48 & 4.08 & 0.14 & 4.61 & 0.19 & 5, 6\\
\hline \hline
    \multicolumn{7}{l}{NOTES: a: A non-member based
upon its radial velocity}\\
\multicolumn{7}{l}{References:}\\
\multicolumn{7}{l}{(1) ID: Stetson \& Harris 1977}\\
\multicolumn{7}{l}{(2) Photometry: Stetson \& Harris 1977}\\
\multicolumn{7}{l}{(3) ID: Alcaino \& Liller 1990}\\
\multicolumn{7}{l}{(4) ID: Harris 1975}\\
\multicolumn{7}{l}{(5) Photometry: Stetson, P.B., private
comm.}\\
\multicolumn{7}{l}{(6) ID: Stetson, P.B., private comm.}\\
\multicolumn{7}{l}{(7) ID: Johnson \& Sadage 1956}\\
\multicolumn{7}{l}{(8) Photometry: Cudworth 1979}

\end{tabular}
\label{table-clusters2}
\end{table*}

\clearpage

\begin{table*}
    \caption{The Calibrating Globular Clusters}
\begin{tabular}{lccccccc}
\hline \hline
 Cluster & $V_{\rm HB}$ & [Fe/H] & $W^\prime_{\rm Rut}$ & $\sigma_{W^\prime_{\rm Rut}}$ & $W^\prime$
 & $\sigma_{W^\prime}$ & N \\
 & (mag) & &(\AA)& (\AA) &(\AA)  & (\AA) & \\
\hline
NGC 1904 = M79  & 16.15 & $-$1.52 & 3.09 & 0.12 & 2.51 & 0.05 & 8 \\
NGC 4590 = M68  & 15.68 & $-$2.30 & 1.59 & 0.08 & 1.32 & 0.03 & 9 \\
NGC 5272 = M3   & 15.65 & $-$1.55 & 3.07 & 0.08 & 2.91 & 0.05 & 8 \\
NGC 6171 = M107 & 15.70 & $-$0.95 & 3.99 & 0.08 & 3.71 & 0.05 & 8 \\
\hline \hline
\end{tabular}
    \label{table-clusters1}
    \end{table*}

\clearpage

\begin{table*}
    \caption{Regressions for \wp to Fe~II Results}
    \begin{tabular}{l|ccccccccc}
    \hline \hline
    &  \multicolumn{9}{c}{[Fe/H] = A + B*\wp + C * $\rm {W^\prime}^2$}\\
    Line Style & A & $\sigma_A$ & B & $\sigma_B$ & C & $\sigma_C$ & m.e.1 & rms & ${\chi_\nu}^2$\\
               & (dex)   &           & (dex/\AA) & & (
               (dex/\AA)$^2$) & & & & \\
    \hline
    Dashed    & $-$2.985 & 0.256&  $+$0.329 & 0.145 & $+$0.029 & 0.019 & 1.27 & 0.14 & 0.04 \\
    Dot-dash & $-$3.340 & 0.055&  $+$0.549 & 0.015 & ... & ... & 1.40 & 0.18 & 4.52 \\
    Solid     & $-$3.078 & 0.058&  $+$0.440 & 0.020 & ... & ... & 1.25 & 0.10 & 0.31 \\
    KI03$^a$  & $-$3.225 & 0.082&  $+$0.537 & 0.024 & ... & ... & ... &...  & \\
    \hline \hline
    \multicolumn{10}{l}{}\\
    \multicolumn{10}{l}{\parbox{5.5in}{\renewcommand{\baselinestretch}{1.}
\small{Dot-dash=linear fit, Dashed=quadratic fit, Solid=linear fit to GGCs only\\
    $^a$: Indicates the fit from KI03 based on Fe~II abundances}}}\\
    \end{tabular}
    \label{fe_fits}
\end{table*}

\clearpage

\begin{table*}
    \caption{Abundances and Reduced Equivalent Widths for Clusters used in the Ca~II
    $\rightarrow$ [Ca/H] Calibration from Bosler, 2004}
    \begin{tabular}{l|ll|lcc|ccccc}
    \hline \hline
    \scriptsize{Group} & NGC & Alt.  & \wp & \sigwp & \# & [Fe/H]
    &  $\rm \sigma_{[Fe/H]}$ & $\rm [Ca/H]$ & $\rm \sigma_{[Ca/H]}$ & \#\\
    ~~\# & ~~\# & ID & (\AA) & (\AA) & of Stars & & & & & of Stars\\
    \hline
    & 104 & 47 Tuc       & 4.53$^b$ & 0.07 & 31 & $-$0.64 & 0.04 & $-$0.50 & 0.04 & 12\\
    & 288 &              & 3.86     & 0.08 & 10 & $-$1.53 & 0.06 & $-$1.11 & 0.13 & 2\\
    & 1904 & M79         & 3.14$^b$ & 0.14 &  9 & $-$1.52 & 0.04 & $-$1.24 & 0.07 & 6\\
    & 4590 & M68         & 1.59$^b$ & 0.08 & 19 & $-$2.30 & 0.04 & $-$2.01 & 0.05 & 7\\
    1 & 5272 & M3        & 3.10     & 0.07 & 9  & $-$1.58 & 0.04 & $-$1.31 & 0.06 & 6\\
    & 5904 & M5          & 3.73$^b$ & 0.12 & 43 & $-$1.38 & 0.06 & $-$0.99 & 0.09 & 4\\
    & 6341 & M92         & 1.77     & 0.08 & 6  & $-$2.43 & 0.04 & $-$2.04 & 0.06 & 7\\
    & 6638  &            & 4.31$^b$ & 0.12 & 12 & $-$1.26 & 0.05 & $-$0.98 & 0.08 & 4 \\
    & 7089 & M2          & 3.26     & 0.07 & 11 & $-$1.72 & 0.08 & $-$1.36 & 0.10 & 4\\
    & 7099 & M30         & 1.89     & 0.09 & 5  & $-$2.22 & 0.04 & $-$1.83 & 0.06 & 5\\
    \hline
    & 6121 & M4          & 3.87$^b$ & 0.07 & 33 & $-$1.37 & 0.11 & $-$1.11& 0.11 &  24 \\
    & 6254 & M10         & 3.42$^b$ & 0.09 & 16 & $-$1.67 & 0.11 & $-$1.38& 0.09 &  9 \\
    & 6397 &             & 2.17$^b$ & 0.08 & 19 & $-$2.11 & 0.11 & $-$1.91& 0.11 &  6 \\
2$^a$ & 6752 &           & 3.41$^b$ & 0.07 & 14 & $-$1.72 & 0.12 & $-$1.46& 0.11 &  8 \\
    & 6838 & M71         & 4.64$^b$ & 0.17 & 11 & $-$1.09 & 0.11 & $-$0.66& 0.09 &  10 \\
    & 7078 & M15         & 1.56$^b$ & 0.12 & 6  & $-$2.45 & 0.11 & $-$2.16& 0.09 &  7 \\
    \hline
    & 2141 &             & 5.41     & 0.09 & 10 & $-$0.46 & 0.11 & $-$0.42 & 0.11 &  5\\
    & 2420 &             & 5.21     & 0.10 & 6  & $-$0.33 & 0.13 & $-$0.41 & 0.14 &  3\\
    3 & 2682 & M67       & 5.62     & 0.10 & 8  & $+$0.03 & 0.10 & $-$0.08 & 0.09 &  6\\
    & 6791 &             & 6.50     & 0.11 & 8  & $+$0.07 & 0.29 & $+$0.02 & 0.24 &  2\\
    & 7789 &             & 5.32     & 0.10 & 8  & $-$0.37 & 0.11 & $-$0.41 & 0.12 &  3\\
\hline \hline \multicolumn{10}{l}{}\\
\multicolumn{11}{l}{\parbox{5.5in}{\renewcommand{\baselinestretch}{1.}
\small{$^a$: The [Ca/H] values for Group 2 clusters have been
derived using the [Ca/Fe] value found in the literature: Ivans
\etal (1999) for M4, Kraft \etal (1995) for M10, Castillo \etal
(2000) for N6397, Cavallo \etal (2004) for N6752, Ramirez \etal
(2002) for M71 and Sneden \etal (1997) for M15. Since these values
have not been transformed to the scale of this work, the calcium
values are listed here but were not used in the final [Ca/H]
calibration of the Ca~II triplet.}}}\\
\multicolumn{11}{l}{\parbox{5.5in}{\renewcommand{\baselinestretch}{1.}
\small{$^b$:~\wp and $\sigma_{W^\prime}$ come from R97a.}}}\\
    \end{tabular}
    \label{wp_abund}
    \end{table*}

\clearpage

%\begin{landscape}
%Leo I LRIS data: 02/08/2002 and 03/23/2003
% Table Leo I
\onecolumn

\begin{longtable}{ccccccc}
\caption[]{Parameters for Stars in the Leo I dSph}\\
    \hline \hline
ID & V & {B--V} & RA & DEC & SNR &$v_{\rm helio}$ \\
 & (mag)& (mag) & (J2000) & (J2000) &  & (km/sec)\\
\hline  \endfirsthead
    \caption[]{Leo I \it{continued}}\\
    \hline \hline
ID & V & {B--V} & RA & DEC & SNR &$v_{\rm helio}$ \\
 & (mag)& (mag) & (J2000) & (J2000) &  & (km/sec)\\
\hline  \endhead
    \hline \hline
    \multicolumn{4}{r}{\it{continued on next page}}
    \endfoot
    \endlastfoot
 2195 & 19.78 & 1.45 & 10:7:59.14 & 12:17:17.6 & 18 & 303.0\\
 2405 & 20.40 & 1.14 & 10:8:01.21 & 12:16:14.7 & 11 & 280.9\\
 2488 & 19.92 & 1.33 & 10:8:01.89 & 12:17:55.6 & 26 & 287.1\\
 2557 & 20.13 & 1.32 & 10:8:02.49 & 12:16:19.8 & 14 & 278.8\\
 2655 & 20.30 & 1.39 & 10:8:03.24 & 12:19:11.0 & 17 & 282.7\\
 2767 & 20.22 & 1.22 & 10:8:04.00 & 12:19:41.6 & 16 & 297.6\\
 2907 & 20.20 & 1.20 & 10:8:04.91 & 12:17:53.3 & 17 & 279.1\\
 3135 & 20.20 & 1.12 & 10:8:06.28 & 12:17:45.1 & 19 & 270.6\\
 3499 & 20.15 & 1.33 & 10:8:08.02 & 12:19:24.3 & 16 & 281.9\\
 3994 & 19.88 & 1.31 & 10:8:09.96 & 12:16:25.4 & 17 & 287.9\\
 4173 & 20.08 & 1.12 & 10:8:10.59 & 12:17:07.9 & 19 & 257.6\\
 4690 & 20.16 & 1.36 & 10:8:12.12 & 12:17:12.4 & 15 & 270.3\\
 5496 & 19.89 & 1.01 & 10:8:14.19 & 12:17:36.1 & 18 & 302.9\\
 6065 & 19.81 & 1.47 & 10:8:15.41 & 12:19:01.4 & 34 & 279.2\\
 6119 & 20.16 & 1.29 & 10:8:15.52 & 12:16:41.8 & 16 & 300.0\\
 6372 & 20.30 & 1.09 & 10:8:16.08 & 12:19:22.7 & 23 & 265.6\\
 6581 & 20.25 & 1.10 & 10:8:16.48 & 12:15:51.4 & 15 & 296.1\\
 6849 & 19.88 & 0.95 & 10:8:16.97 & 12:18:28.3 & 26 & 293.0\\
 7239 & 20.21 & 1.13 & 10:8:17.61 & 12:16:30.4 & 22 & 266.4\\
 7548 & 19.92 & 1.23 & 10:8:18.14 & 12:17:46.2 & 25 & 270.3\\
 7752 & 20.23 & 1.01 & 10:8:18.47 & 12:19:51.1 & 12 & 241.5\\
 7975 & 20.26 & 1.13 & 10:8:18.82 & 12:19:50.9 & 26 & 269.9\\
 7984 & 20.15 & 1.18 & 10:8:18.83 & 12:19:04.5 & 13 & 274.1\\
 8203 & 19.91 & 1.32 & 10:8:19.16 & 12:20:48.7 & 21 & 281.9\\
 8391 & 20.26 & 1.01 & 10:8:19.44 & 12:16:34.2 & 13 & 282.1\\
 8409 & 20.11 & 1.21 & 10:8:19.47 & 12:19:09.7 & 14 & 273.8\\
 8608 & 19.76 & 1.27 & 10:8:19.75 & 12:16:55.0 & 33 & 298.8\\
 8635 & 20.32 & 1.08 & 10:8:19.79 & 12:16:34.2 & 13 & 282.1\\
 8885 & 19.65 & 1.13 & 10:8:20.13 & 12:17:02.8 & 27 & 288.0\\
 8893 & 20.18 & 0.94 & 10:8:20.15 & 12:17:47.0 & 16 & 286.1\\
 8937 & 19.78 & 1.08 & 10:8:20.21 & 12:17:56.0 & 14 & 266.8\\
 9099 & 20.33 & 1.07 & 10:8:20.45 & 12:16:01.7 & 21 & 283.4\\
 9187 & 20.24 & 0.98 & 10:8:20.55 & 12:15:47.2 & 14 & 263.7\\
 9241 & 20.33 & 1.19 & 10:8:20.63 & 12:19:21.0 & 14 & 276.6\\
 9683 & 19.93 & 1.36 & 10:8:21.23 & 12:17:41.3 & 15 & 283.5\\
 9764 & 19.83 & 1.35 & 10:8:21.34 & 12:19:52.6 & 12 & 284.3\\
 9782 & 19.56 & 1.17 & 10:8:21.36 & 12:16:22.8 & 23 & 279.7\\
 9915 & 19.68 & 1.33 & 10:8:21.55 & 12:19:12.9 & 35 & 283.0\\
10190 & 20.09 & 1.30 & 10:8:21.87 & 12:19:33.2 & 17 & 271.2\\
10439 & 20.01 & 1.21 & 10:8:22.22 & 12:21:39.5 & 18 & 279.1\\
10670 & 20.12 & 1.21 & 10:8:22.50 & 12:22:40.1 & 26 & 275.6\\
10789 & 19.91 & 1.28 & 10:8:22.65 & 12:17:21.5 & 18 & 252.8\\
10914 & 19.80 & 1.19 & 10:8:22.80 & 12:17:59.9 & 29 & 290.3\\
11712 & 20.43 & 1.10 & 10:8:23.77 & 12:15:43.9 & 18 & 295.0\\
11758 & 19.63 & 1.20 & 10:8:23.83 & 12:17:10.7 & 36 & 273.6\\
11808 & 19.69 & 1.28 & 10:8:23.89 & 12:17:37.3 & 23 & 290.0\\
12171 & 20.33 & 0.89 & 10:8:24.35 & 12:18:27.4 & 11 & 296.4\\
12351 & 19.78 & 1.21 & 10:8:24.54 & 12:17:42.1 & 22 & 288.2\\
12515 & 19.87 & 1.40 & 10:8:24.71 & 12:17:24.2 & 28 & 276.0\\
12581 & 20.24 & 1.17 & 10:8:24.81 & 12:15:54.6 & 19 & 269.1\\
13094 & 20.02 & 1.37 & 10:8:25.39 & 12:16:53.3 & 18 & 284.1\\
13137 & 20.01 & 1.29 & 10:8:25.43 & 12:17:16.6 & 26 & 245.9\\
13158 & 19.88 & 1.24 & 10:8:25.46 & 12:16:36.3 & 25 & 288.6\\
13415 & 20.13 & 1.48 & 10:8:25.74 & 12:21:01.9 & 13 & 285.2\\
14095 & 19.81 & 1.34 & 10:8:26.49 & 12:19:11.2 & 34 & 279.6\\
14307 & 20.53 & 1.19 & 10:8:26.71 & 12:16:36.4 & 16 & 275.7\\
14738 & 19.92 & 0.84 & 10:8:27.16 & 12:17:56.8 & 14 & 284.6\\
14927 & 19.73 & 1.25 & 10:8:27.36 & 12:16:17.6 & 23 & 282.6\\
16048 & 19.79 & 1.13 & 10:8:28.56 & 12:19:08.0 & 14 & 295.2\\
16088 & 20.12 & 1.20 & 10:8:28.59 & 12:16:20.6 & 25 & 280.8\\
16219 & 19.64 & 1.19 & 10:8:28.76 & 12:17:55.1 & 25 & 284.0\\
16248 & 20.33 & 1.05 & 10:8:28.79 & 12:20:33.6 & 16 & 272.8\\
16334 & 19.90 & 1.21 & 10:8:28.89 & 12:19:27.8 & 18 & 291.6\\
16635 & 19.78 & 1.25 & 10:8:29.20 & 12:16:12.2 & 33 & 283.3\\
17009 & 19.82 & 1.36 & 10:8:29.60 & 12:18:07.5 & 32 & 289.7\\
17226 & 20.32 & 1.13 & 10:8:29.86 & 12:19:14.6 & 22 & 298.1\\
17437 & 20.12 & 0.96 & 10:8:30.08 & 12:18:12.6 & 15 & 265.3\\
17620 & 20.40 & 0.98 & 10:8:30.29 & 12:17:13.5 & 12 & 300.6\\
17854 & 19.74 & 1.11 & 10:8:30.56 & 12:16:27.5 & 34 & 293.8\\
18030 & 19.56 & 1.12 & 10:8:30.76 & 12:17:29.8 & 26 & 285.6\\
18214 & 19.77 & 1.35 & 10:8:30.97 & 12:16:40.6 & 27 & 286.4\\
18315 & 19.69 & 1.28 & 10:8:31.09 & 12:18:35.8 & 30 & 270.8\\
18323 & 20.08 & 0.94 & 10:8:31.09 & 12:18:02.6 & 21 & 298.4\\
18509 & 20.03 & 1.19 & 10:8:31.31 & 12:20:21.9 & 19 & 297.2\\
18588 & 19.87 & 1.46 & 10:8:31.40 & 12:15:55.8 & 26 & 288.6\\
18676 & 20.44 & 1.14 & 10:8:31.50 & 12:21:49.1 & 13 & 274.4\\
18765 & 20.03 & 1.34 & 10:8:31.58 & 12:17:26.4 & 24 & 287.0\\
18801 & 19.74 & 1.29 & 10:8:31.62 & 12:18:50.3 & 18 & 287.5\\
18812 & 20.13 & 1.13 & 10:8:31.64 & 12:16:52.4 & 16 & 293.0\\
18880 & 20.05 & 1.16 & 10:8:31.71 & 12:18:38.6 & 11 & 271.7\\
19040 & 20.08 & 1.20 & 10:8:31.88 & 12:22:30.9 & 14 & 291.4\\
19080 & 19.71 & 1.18 & 10:8:31.93 & 12:18:41.9 & 31 & 275.1\\
19262 & 20.11 & 1.22 & 10:8:32.14 & 12:18:38.5 & 11 & 293.9\\
19667 & 19.73 & 1.22 & 10:8:32.62 & 12:18:09.0 & 31 & 265.8\\
19881 & 20.02 & 1.15 & 10:8:32.87 & 12:17:35.2 & 27 & 290.1\\
20470 & 19.75 & 1.18 & 10:8:33.57 & 12:18:25.7 & 28 & 279.5\\
20882 & 19.96 & 1.33 & 10:8:34.06 & 12:23:16.7 & 15 & 275.8\\
21026 & 20.06 & 1.23 & 10:8:34.24 & 12:16:11.4 & 16 & 292.1\\
21192 & 20.28 & 1.19 & 10:8:34.45 & 12:21:23.0 & 14 & 286.8\\
21196 & 20.54 & 1.13 & 10:8:34.46 & 12:20:06.5 & 11 & 302.1\\
21325 & 19.70 & 1.33 & 10:8:34.64 & 12:18:56.2 & 26 & 297.3\\
21348 & 20.29 & 1.11 & 10:8:34.67 & 12:17:16.7 & 33 & 270.4\\
22001 & 20.35 & 1.20 & 10:8:35.55 & 12:16:59.1 & 23 & 279.1\\
22513 & 20.07 & 1.30 & 10:8:36.25 & 12:19:12.1 & 26 & 298.3\\
22788 & 19.73 & 1.34 & 10:8:36.65 & 12:17:50.6 & 11 & 271.5\\
23074 & 20.08 & 1.25 & 10:8:37.06 & 12:16:56.3 & 23 & 282.2\\
23518 & 20.35 & 0.99 & 10:8:37.68 & 12:19:49.6 & 18 & 267.9\\
24095 & 20.26 & 0.99 & 10:8:38.65 & 12:18:51.8 & 17 & 263.8\\
25113 & 19.70 & 1.38 & 10:8:40.45 & 12:16:37.9 & 26 & 287.1\\
25440 & 20.38 & 0.90 & 10:8:41.10 & 12:17:36.5 & 17 & 292.1\\
25820 & 20.20 & 1.23 & 10:8:41.92 & 12:18:39.5 & 21 & 251.9\\
 \hline \hline
\label{table-leo1}

    \end{longtable}

%\end{landscape}

\clearpage

%\end{landscape}

\clearpage

%Leo II LRIS data: 02/08/2002 and 03/23/2003
% Table Leo II
\begin{longtable}{ccccccc}
\caption[]{Parameters for Stars in the Leo I dSph}\\
    \hline \hline
ID & V & {B--V} & RA & DEC & SNR &$v_{\rm helio}$ \\
 & (mag)& (mag) & (J2000) & (J2000) &  & (km/sec)\\
\hline  \endfirsthead
    \caption[]{Leo I \it{continued}}\\
    \hline \hline
ID & V & {B--V} & RA & DEC & SNR &$v_{\rm helio}$ \\
 & (mag)& (mag) & (J2000) & (J2000) &  & (km/sec)\\
\hline  \endhead
    \hline \hline
    \multicolumn{4}{r}{\it{continued on next page}}
    \endfoot
    \endlastfoot
156 & 18.58 & 1.79 & 11:13:20.9 & 22:08:22.9 & 19 & 103.2\\
166 & 18.87 & 1.17 & 11:13:38.9 & 22:10:18.2 & 38 & 92.7\\
180 & 19.08 & 1.47 & 11:13:23.2 & 22:10:55.9 & 42 & 84.7\\
195 & 19.08 & 1.08 & 11:13:33.2 & 22:08:12.5 & 36 & 88.2\\
209 & 19.16 & 1.42 & 11:13:31.1 & 22:06:30.4 & 42 & 98.6\\
230 & 19.23 & 1.41 & 11:13:34.8 & 22:09:42.4 & 27 & 108.0\\
233 & 19.22 & 1.34 & 11:13:24.5 & 22:06:13.5 & 18 & 75.6\\
234 & 19.26 & 1.21 & 11:13:20.8 & 22:08:34.4 & 21 & 75.1\\
235 & 19.24 & 1.44 & 11:13:23.4 & 22:08:54.3 & 22 & 94.6\\
236 & 19.21 & 1.37 & 11:13:40.5 & 22:12:18.4 & 37 & 101.1\\
237 & 19.26 & 1.04 & 11:13:39.8 & 22:10:23.1 & 30 & 100.0\\
248 & 19.31 & 1.20 & 11:13:27.7 & 22:10:13.0 & 29 & 98.7\\
249 & 19.32 & 1.16 & 11:13:19.3 & 22:07:41.6 & 22 & 106.8\\
254 & 19.35 & 1.34 & 11:13:32.1 & 22:07:29.8 & 34 & 84.8\\
255 & 19.30 & 1.35 & 11:13:34.5 & 22:11:42.5 & 38 & 86.7\\
256 & 19.34 & 1.23 & 11:13:27.7 & 22:10:39.7 & 30 & 99.7\\
258 & 19.35 & 1.35 & 11:13:19.8 & 22:09:20.3 & 37 & 91.1\\
259 & 19.35 & 1.37 & 11:13:30.8 & 22:10:51.0 & 18 & 100.2\\
260 & 19.32 & 1.29 & 11:13:26.6 & 22:07:26.3 & 33 & 113.4\\
271 & 19.37 & 1.37 & 11:13:19.9 & 22:09:45.9 & 40 & 74.2\\
277 & 19.39 & 1.26 & 11:13:37.8 & 22:09:27.2 & 22 & 85.2\\
280 & 19.45 & 1.12 & 11:13:23.8 & 22:05:32.8 & 28 & 90.7\\
281 & 19.42 & 1.09 & 11:13:32.7 & 22:07:04.7 & 35 & 82.1\\
282 & 19.38 & 1.48 & 11:13:39.5 & 22:09:16.8 & 34 & 95.3\\
285 & 19.43 & 1.31 & 11:13:33.9 & 22:06:08.2 & 34 & 79.3\\
293 & 19.37 & 1.29 & 11:13:25.7 & 22:06:37.8 & 37 & 60.4\\
296 & 19.46 & 1.05 & 11:13:21.4 & 22:10:37.7 & 35 & 64.6\\
302 & 19.46 & 1.17 & 11:13:30.9 & 22:08:11.5 & 30 & 117.2\\
304 & 19.47 & 1.25 & 11:13:30.3 & 22:06:57.6 & 35 & 82.9\\
320 & 19.47 & 1.30 & 11:13:29.7 & 22:06:46.5 & 18 & 87.3\\
331 & 19.56 & 1.12 & 11:13:37.1 & 22:09:33.2 & 33 & 101.7\\
333 & 19.56 & 1.20 & 11:13:26.5 & 22:11:07.4 & 33 & 76.4\\
336 & 19.56 & 1.18 & 11:13:23.3 & 22:10:05.2 & 37 & 91.0\\
341 & 19.59 & 1.24 & 11:13:29.6 & 22:08:57.7 & 32 & 87.4\\
351 & 19.59 & 1.25 & 11:13:31.0 & 22:11:21.9 & 39 & 79.1\\
352 & 19.55 & 1.36 & 11:13:23.1 & 22:10:22.6 & 17 & 92.4\\
370 & 19.64 & 0.93 & 11:13:33.3 & 22:09:44.3 & 21 & 88.2\\
377 & 19.65 & 1.19 & 11:13:29.6 & 22:07:16.1 & 27 & 84.3\\
379 & 19.67 & 1.29 & 11:13:31.2 & 22:08:24.6 & 33 & 72.5\\
392 & 19.70 & 1.13 & 11:13:28.8 & 22:08:47.0 & 12 & 106.0\\
395 & 19.73 & 1.14 & 11:13:33.1 & 22:07:37.6 & 23 & 91.7\\
396 & 19.71 & 0.95 & 11:13:27.9 & 22:08:15.4 & 18 & 85.2\\
397 & 19.73 & 1.17 & 11:13:34.5 & 22:08:09.8 & 19 & 80.6\\
402 & 19.74 & 1.19 & 11:13:31.0 & 22:08:44.6 & 24 & 86.5\\
409 & 19.74 & 1.06 & 11:13:27.9 & 22:08:29.5 & 11 & 78.5\\
420 & 19.82 & 1.11 & 11:13:30.9 & 22:09:32.6 & 28 & 79.7\\
422 & 19.79 & 1.10 & 11:13:16.6 & 22:08:58.9 & 21 & 80.0\\
429 & 19.73 & 1.22 & 11:13:25.6 & 22:07:44.1 & 19 & 79.4\\
430 & 19.79 & 1.18 & 11:13:28.5 & 22:08:43.2 & 19 & 79.9\\
432 & 19.83 & 0.97 & 11:13:15.5 & 22:09:00.9 & 14 & 82.4\\
434 & 19.81 & 1.14 & 11:13:31.7 & 22:07:46.8 & 21 & 67.0\\
435 & 19.78 & 1.20 & 11:13:36.0 & 22:12:20.8 & 21 & 99.5\\
441 & 19.83 & 1.01 & 11:13:31.6 & 22:08:34.2 & 20 & 88.4\\
471 & 19.88 & 0.95 & 11:13:38.5 & 22:09:47.3 & 12 & 90.4\\
493 & 19.96 & 1.04 & 11:13:35.9 & 22:09:27.7 & 16 & 99.5\\
495 & 19.95 & 1.14 & 11:13:32.5 & 22:11:22.8 & 22 & 92.7\\
508 & 19.97 & 1.11 & 11:13:23.7 & 22:07:07.2 & 22 & 86.9\\
524 & 20.04 & 1.18 & 11:13:35.1 & 22:11:34.7 & 17 & 95.6\\
526 & 20.06 & 1.12 & 11:13:36.2 & 22:10:18.6 & 16 & 86.9\\
533 & 20.06 & 1.03 & 11:13:26.8 & 22:08:24.8 & 15 & 70.7\\
541 & 20.10 & 1.08 & 11:13:23.7 & 22:09:12.6 & 15 & 60.7\\
549 & 20.13 & 1.07 & 11:13:18.2 & 22:08:29.3 & 15 & 88.5\\
552 & 20.15 & 1.01 & 11:13:32.8 & 22:10:53.3 & 22 & 103.5\\
556 & 20.14 & 1.00 & 11:13:38.0 & 22:11:17.4 & 13 & 105.4\\
558 & 20.20 & 0.95 & 11:13:23.3 & 22:05:23.3 & 24 & 57.9\\
569 & 20.15 & 1.06 & 11:13:23.4 & 22:06:18.1 & 22 & 85.3\\
570 & 20.17 & 1.09 & 11:13:34.7 & 22:11:01.9 & 12 & 111.8\\
573 & 20.18 & 1.16 & 11:13:36.2 & 22:08:51.2 & 24 & 92.6\\
576 & 20.20 & 1.04 & 11:13:30.5 & 22:08:24.9 & 16 & 86.0\\
584 & 20.20 & 1.04 & 11:13:32.3 & 22:06:55.9 & 17 & 78.6\\
588 & 20.22 & 1.00 & 11:13:33.8 & 22:10:03.3 & 15 & 83.6\\
604 & 20.24 & 1.05 & 11:13:37.9 & 22:08:13.3 & 14 & 74.2\\
614 & 20.30 & 0.99 & 11:13:34.5 & 22:11:05.5 & 18 & 86.9\\
623 & 20.26 & 1.22 & 11:13:26.1 & 22:06:48.2 & 17 & 83.0\\
 \hline \hline
\label{table-leo2}

    \end{longtable}

%\end{landscape}

\clearpage

\begin{longtable}{ccccccccc}
\caption[]{Spectroscopic Data for Stars in the Leo I dSph}\\
    \hline \hline
ID & \wp$^a$ & \sigwp & [Fe/H] & $\rm \sigma_{[Fe/H]}$ & $\rm
\sigma_{[Fe/H]}$ & [Ca/H] & $\rm
\sigma_{[Ca/H]}$ & $\rm \sigma_{[Ca/H]}$ \\
 & (\AA) & (\AA)& CG97 & Total & Random & & Total & Random \\
\hline  \endfirsthead
    \caption[]{Leo I \it{continued}}\\
    \hline \hline
ID & \wp$^a$ & \sigwp & [Fe/H] & $\rm \sigma_{[Fe/H]}$ & $\rm
\sigma_{[Fe/H]}$ & [Ca/H] & $\rm
\sigma_{[Ca/H]}$ & $\rm \sigma_{[Ca/H]}$ \\
 & (\AA) & (\AA)& CG97 & Total & Random & & Total & Random \\
\hline  \endhead
    \hline \hline
    \multicolumn{4}{r}{\it{continued on next page}}
    \endfoot
    \endlastfoot
2195 & 3.76 & 0.20 & $-$1.08 & 0.16 & 0.08 & $-$1.01 & 0.15 & 0.09\\
2405 & 3.17 & 0.30 & $-$1.33 & 0.21 & 0.11 & $-$1.29 & 0.22 & 0.14\\
2488 & 2.92 & 0.19 & $-$1.43 & 0.15 & 0.06 & $-$1.41 & 0.15 & 0.09\\
2557 & 2.90 & 0.23 & $-$1.44 & 0.17 & 0.08 & $-$1.42 & 0.17 & 0.11\\
2655 & 1.99 & 0.21 & $-$1.82 & 0.16 & 0.07 & $-$1.84 & 0.16 & 0.10\\
2767 & 2.50 & 0.24 & $-$1.61 & 0.18 & 0.09 & $-$1.60 & 0.18 & 0.11\\
2907 & 2.84 & 0.22 & $-$1.47 & 0.17 & 0.08 & $-$1.44 & 0.16 & 0.10\\
3135 & 3.03 & 0.24 & $-$1.39 & 0.18 & 0.09 & $-$1.35 & 0.18 & 0.11\\
3499 & 3.79 & 0.23 & $-$1.07 & 0.17 & 0.09 & $-$1.00 & 0.17 & 0.11\\
3994 & 2.59 & 0.19 & $-$1.57 & 0.15 & 0.06 & $-$1.56 & 0.15 & 0.09\\
4173 & 2.77 & 0.21 & $-$1.50 & 0.16 & 0.07 & $-$1.48 & 0.16 & 0.10\\
4690 & 3.81 & 0.26 & $-$1.06 & 0.19 & 0.10 & $-$0.99 & 0.19 & 0.12\\
5496 & 2.50 & 0.22 & $-$1.61 & 0.17 & 0.08 & $-$1.60 & 0.16 & 0.10\\
6065 & 3.50 & 0.19 & $-$1.19 & 0.15 & 0.07 & $-$1.13 & 0.15 & 0.09\\
6119 & 2.56 & 0.22 & $-$1.58 & 0.17 & 0.08 & $-$1.57 & 0.16 & 0.10\\
6372 & 1.86 & 0.20 & $-$1.88 & 0.15 & 0.06 & $-$1.90 & 0.15 & 0.09\\
6581 & 3.66 & 0.36 & $-$1.12 & 0.25 & 0.15 & $-$1.06 & 0.26 & 0.17\\
6849 & 2.34 & 0.23 & $-$1.68 & 0.17 & 0.08 & $-$1.68 & 0.17 & 0.11\\
7239 & 2.58 & 0.19 & $-$1.57 & 0.15 & 0.06 & $-$1.57 & 0.15 & 0.09\\
7548 & 4.04 & 0.23 & $-$0.96 & 0.17 & 0.09 & $-$0.88 & 0.17 & 0.11\\
7752 & 1.77 & 0.28 & $-$1.92 & 0.20 & 0.10 & $-$1.95 & 0.20 & 0.13\\
7975 & 3.64 & 0.21 & $-$1.13 & 0.16 & 0.08 & $-$1.07 & 0.16 & 0.10\\
7984 & 3.85 & 0.20 & $-$1.04 & 0.16 & 0.07 & $-$0.97 & 0.15 & 0.09\\
8203 & 3.72 & 0.20 & $-$1.10 & 0.16 & 0.07 & $-$1.03 & 0.15 & 0.09\\
8391 & 2.92 & 0.35 & $-$1.41 & 0.15 & 0.12 & $-$1.41 & 0.25 & 0.16\\
8409 & 3.09 & 0.24 & $-$1.36 & 0.18 & 0.09 & $-$1.33 & 0.18 & 0.11\\
8608 & 2.83 & 0.24 & $-$1.47 & 0.18 & 0.09 & $-$1.45 & 0.18 & 0.11\\
8635 & 3.13 & 0.23 & $-$1.13 & 0.22 & 0.11 & $-$1.31 & 0.17 & 0.11\\
8885 & 2.22 & 0.21 & $-$1.73 & 0.16 & 0.07 & $-$1.73 & 0.16 & 0.10\\
8893 & 3.58 & 0.26 & $-$1.16 & 0.19 & 0.10 & $-$1.10 & 0.19 & 0.12\\
8937 & 4.21 & 0.28 & $-$0.89 & 0.20 & 0.11 & $-$0.80 & 0.20 & 0.13\\
9099 & 3.40 & 0.20 & $-$1.23 & 0.16 & 0.07 & $-$1.18 & 0.15 & 0.09\\
9187 & 3.28 & 0.28 & $-$1.28 & 0.20 & 0.11 & $-$1.24 & 0.20 & 0.13\\
9241 & 2.84 & 0.24 & $-$1.47 & 0.18 & 0.09 & $-$1.44 & 0.18 & 0.11\\
9683 & 3.01 & 0.22 & $-$1.40 & 0.17 & 0.08 & $-$1.36 & 0.17 & 0.10\\
9764 & 2.40 & 0.29 & $-$1.65 & 0.21 & 0.11 & $-$1.65 & 0.21 & 0.14\\
9782 & 2.25 & 0.18 & $-$1.71 & 0.15 & 0.05 & $-$1.72 & 0.14 & 0.08\\
9915 & 3.95 & 0.21 & $-$1.00 & 0.16 & 0.08 & $-$0.92 & 0.16 & 0.10\\
10190 & 2.70 & 0.21 & $-$1.53 & 0.16 & 0.07 & $-$1.51 & 0.16 & 0.10\\
10439 & 3.27 & 0.19 & $-$1.28 & 0.15 & 0.07 & $-$1.24 & 0.15 & 0.09\\
10670 & 3.40 & 0.23 & $-$1.23 & 0.17 & 0.08 & $-$1.18 & 0.17 & 0.11\\
10789 & 3.36 & 0.22 & $-$1.25 & 0.17 & 0.08 & $-$1.20 & 0.17 & 0.10\\
10914 & 3.09 & 0.21 & $-$1.36 & 0.16 & 0.07 & $-$1.33 & 0.16 & 0.10\\
11712 & 2.25 & 0.26 & $-$1.72 & 0.19 & 0.09 & $-$1.72 & 0.19 & 0.12\\
11758 & 2.70 & 0.19 & $-$1.53 & 0.15 & 0.06 & $-$1.51 & 0.15 & 0.09\\
11808 & 3.45 & 0.20 & $-$1.21 & 0.16 & 0.07 & $-$1.16 & 0.15 & 0.09\\
12171 & 3.24 & 0.30 & $-$1.30 & 0.21 & 0.12 & $-$1.26 & 0.22 & 0.14\\
12351 & 3.00 & 0.22 & $-$1.40 & 0.17 & 0.08 & $-$1.37 & 0.17 & 0.10\\
12515 & 2.80 & 0.24 & $-$1.49 & 0.18 & 0.09 & $-$1.46 & 0.18 & 0.11\\
12581 & 3.02 & 0.20 & $-$1.39 & 0.16 & 0.07 & $-$1.36 & 0.15 & 0.09\\
13094 & 2.87 & 0.21 & $-$1.45 & 0.16 & 0.07 & $-$1.43 & 0.16 & 0.10\\
13137 & 3.21 & 0.22 & $-$1.31 & 0.17 & 0.08 & $-$1.27 & 0.17 & 0.10\\
13158 & 3.78 & 0.21 & $-$1.07 & 0.16 & 0.08 & $-$1.00 & 0.16 & 0.10\\
13415 & 3.04 & 0.21 & $-$1.38 & 0.16 & 0.07 & $-$1.35 & 0.16 & 0.10\\
14095 & 3.19 & 0.17 & $-$1.32 & 0.14 & 0.06 & $-$1.28 & 0.13 & 0.08\\
14307 & 4.63 & 0.26 & $-$0.72 & 0.19 & 0.11 & $-$0.60 & 0.19 & 0.12\\
14738 & 3.16 & 0.24 & $-$1.33 & 0.18 & 0.09 & $-$1.29 & 0.18 & 0.11\\
14927 & 3.25 & 0.20 & $-$1.29 & 0.16 & 0.07 & $-$1.25 & 0.15 & 0.09\\
16048 & 2.35 & 0.24 & $-$1.67 & 0.18 & 0.08 & $-$1.67 & 0.18 & 0.11\\
16088 & 2.97 & 0.21 & $-$1.41 & 0.16 & 0.07 & $-$1.38 & 0.16 & 0.10\\
16219 & 3.11 & 0.21 & $-$1.35 & 0.16 & 0.07 & $-$1.32 & 0.16 & 0.10\\
16248 & 4.42 & 0.29 & $-$0.80 & 0.21 & 0.12 & $-$0.70 & 0.21 & 0.14\\
16334 & 3.42 & 0.23 & $-$1.22 & 0.17 & 0.08 & $-$1.17 & 0.17 & 0.11\\
16635 & 3.14 & 0.21 & $-$1.34 & 0.16 & 0.07 & $-$1.30 & 0.16 & 0.10\\
17009 & 2.78 & 0.21 & $-$1.49 & 0.16 & 0.07 & $-$1.47 & 0.16 & 0.10\\
17226 & 4.00 & 0.23 & $-$0.98 & 0.17 & 0.09 & $-$0.90 & 0.17 & 0.11\\
17437 & 2.94 & 0.26 & $-$1.43 & 0.19 & 0.10 & $-$1.40 & 0.19 & 0.12\\
17620 & 3.53 & 0.26 & $-$1.18 & 0.19 & 0.10 & $-$1.12 & 0.19 & 0.12\\
17854 & 0.83 & 0.19 & $-$2.31 & 0.15 & 0.05 & $-$2.39 & 0.14 & 0.09\\
18030 & 3.89 & 0.19 & $-$1.03 & 0.15 & 0.07 & $-$0.95 & 0.15 & 0.09\\
18214 & 2.10 & 0.20 & $-$1.78 & 0.16 & 0.06 & $-$1.79 & 0.15 & 0.09\\
18315 & 2.93 & 0.17 & $-$1.43 & 0.14 & 0.05 & $-$1.40 & 0.13 & 0.08\\
18323 & 3.04 & 0.22 & $-$1.38 & 0.17 & 0.08 & $-$1.35 & 0.17 & 0.10\\
18509 & 3.88 & 0.21 & $-$1.03 & 0.16 & 0.08 & $-$0.95 & 0.16 & 0.10\\
18588 & 3.45 & 0.20 & $-$1.21 & 0.16 & 0.07 & $-$1.16 & 0.15 & 0.09\\
18676 & 3.65 & 0.27 & $-$1.13 & 0.19 & 0.10 & $-$1.06 & 0.20 & 0.13\\
18765 & 3.52 & 0.20 & $-$1.18 & 0.16 & 0.07 & $-$1.12 & 0.15 & 0.09\\
18801 & 3.22 & 0.22 & $-$1.31 & 0.17 & 0.08 & $-$1.26 & 0.17 & 0.10\\
18812 & 2.32 & 0.23 & $-$1.68 & 0.17 & 0.08 & $-$1.69 & 0.17 & 0.11\\
18880 & 3.21 & 0.37 & $-$1.31 & 0.26 & 0.15 & $-$1.27 & 0.26 & 0.17\\
19040 & 3.14 & 0.26 & $-$1.34 & 0.19 & 0.10 & $-$1.30 & 0.19 & 0.12\\
19080 & 2.09 & 0.18 & $-$1.78 & 0.15 & 0.05 & $-$1.80 & 0.14 & 0.08\\
19262 & 3.76 & 0.23 & $-$1.08 & 0.18 & 0.09 & $-$1.01 & 0.17 & 0.11\\
19667 & 1.90 & 0.17 & $-$1.86 & 0.14 & 0.04 & $-$1.89 & 0.13 & 0.08\\
19881 & 3.41 & 0.22 & $-$1.23 & 0.17 & 0.08 & $-$1.18 & 0.17 & 0.10\\
20470 & 3.50 & 0.22 & $-$1.19 & 0.17 & 0.08 & $-$1.13 & 0.17 & 0.10\\
20882 & 3.08 & 0.26 & $-$1.36 & 0.19 & 0.10 & $-$1.33 & 0.19 & 0.12\\
21026 & 2.78 & 0.21 & $-$1.49 & 0.16 & 0.07 & $-$1.47 & 0.16 & 0.10\\
21192 & 2.73 & 0.22 & $-$1.52 & 0.17 & 0.08 & $-$1.49 & 0.16 & 0.10\\
21196 & 2.63 & 0.28 & $-$1.56 & 0.20 & 0.11 & $-$1.54 & 0.20 & 0.13\\
21325 & 3.15 & 0.20 & $-$1.34 & 0.16 & 0.07 & $-$1.30 & 0.15 & 0.09\\
21348 & 3.91 & 0.17 & $-$1.02 & 0.14 & 0.06 & $-$0.94 & 0.14 & 0.08\\
22001 & 3.56 & 0.19 & $-$1.16 & 0.15 & 0.07 & $-$1.10 & 0.15 & 0.09\\
22513 & 3.00 & 0.19 & $-$1.40 & 0.15 & 0.06 & $-$1.37 & 0.15 & 0.09\\
22788 & 4.11 & 0.32 & $-$0.93 & 0.23 & 0.13 & $-$0.85 & 0.23 & 0.15\\
23074 & 3.05 & 0.17 & $-$1.38 & 0.14 & 0.05 & $-$1.34 & 0.13 & 0.08\\
23518 & 3.40 & 0.20 & $-$1.23 & 0.16 & 0.07 & $-$1.18 & 0.15 & 0.09\\
24095 & 3.09 & 0.20 & $-$1.36 & 0.16 & 0.07 & $-$1.33 & 0.15 & 0.09\\
25113 & 3.56 & 0.19 & $-$1.17 & 0.15 & 0.07 & $-$1.10 & 0.15 & 0.09\\
25440 & 1.55 & 0.25 & $-$2.01 & 0.18 & 0.09 & $-$2.05 & 0.18 & 0.12\\
25820 & 3.40 & 0.20 & $-$1.23 & 0.16 & 0.07 & $-$1.18 & 0.15 & 0.09\\
  \hline  \hline
  \multicolumn{9}{l}{\renewcommand{\baselinestretch}{1.}
\small{$^a$ The \wp values listed here have been scaled to the
work of R97a.}} \label{table-leo1_spec}

 \end{longtable}

%\end{landscape}

\clearpage

\begin{longtable}{ccccccccc}
\caption[]{Spectroscopic Data for Stars in the Leo II dSph}\\
    \hline \hline
ID & \wp$^a$ & \sigwp & [Fe/H] & $\rm \sigma_{[Fe/H]}$ & $\rm
\sigma_{[Fe/H]}$ & [Ca/H] & $\rm
\sigma_{[Ca/H]}$ & $\rm \sigma_{[Ca/H]}$ \\
 & (\AA) & (\AA)& CG97 & Total & Random & & Total & Random \\
\hline  \endfirsthead
    \caption[]{Leo I \it{continued}}\\
    \hline \hline
ID & \wp$^a$ & \sigwp & [Fe/H] & $\rm \sigma_{[Fe/H]}$ & $\rm
\sigma_{[Fe/H]}$ & [Ca/H] & $\rm
\sigma_{[Ca/H]}$ & $\rm \sigma_{[Ca/H]}$ \\
 & (\AA) & (\AA)& CG97 & Total & Random & & Total & Random \\
\hline  \endhead
    \hline \hline
    \multicolumn{4}{r}{\it{continued on next page}}
    \endfoot
    \endlastfoot
156 & 2.87 & 0.19 & $-$1.45 & 0.15 & 0.06 & $-$1.45 & 0.15 & 0.09\\
166 & 2.71 & 0.18 & $-$1.52 & 0.15 & 0.06 & $-$1.53 & 0.14 & 0.08\\
180 & 2.52 & 0.19 & $-$1.60 & 0.15 & 0.06 & $-$1.61 & 0.15 & 0.09\\
195 & 2.53 & 0.19 & $-$1.60 & 0.15 & 0.06 & $-$1.61 & 0.15 & 0.09\\
209 & 2.54 & 0.18 & $-$1.59 & 0.15 & 0.06 & $-$1.60 & 0.14 & 0.08\\
230 & 2.61 & 0.21 & $-$1.56 & 0.16 & 0.07 & $-$1.57 & 0.16 & 0.10\\
233 & 2.49 & 0.19 & $-$1.62 & 0.15 & 0.06 & $-$1.63 & 0.15 & 0.09\\
234 & 2.05 & 0.19 & $-$1.80 & 0.15 & 0.05 & $-$1.83 & 0.15 & 0.09\\
235 & 3.26 & 0.25 & $-$1.29 & 0.18 & 0.09 & $-$1.28 & 0.18 & 0.11\\
236 & 1.81 & 0.19 & $-$1.90 & 0.15 & 0.05 & $-$1.94 & 0.15 & 0.09\\
237 & 0.36 & 0.22 & $-$2.51 & 0.16 & 0.06 & $-$2.59 & 0.16 & 0.10\\
248 & 2.74 & 0.18 & $-$1.51 & 0.14 & 0.05 & $-$1.51 & 0.14 & 0.08\\
249 & 0.77 & 0.25 & $-$2.34 & 0.18 & 0.08 & $-$2.41 & 0.18 & 0.11\\
254 & 2.94 & 0.19 & $-$1.43 & 0.15 & 0.06 & $-$1.42 & 0.15 & 0.09\\
255 & 2.38 & 0.18 & $-$1.66 & 0.14 & 0.05 & $-$1.68 & 0.14 & 0.08\\
256 & 2.68 & 0.20 & $-$1.54 & 0.16 & 0.07 & $-$1.54 & 0.15 & 0.09\\
258 & 2.59 & 0.18 & $-$1.57 & 0.15 & 0.06 & $-$1.58 & 0.14 & 0.08\\
259 & 3.54 & 0.29 & $-$1.17 & 0.21 & 0.11 & $-$1.15 & 0.21 & 0.13\\
260 & 2.60 & 0.18 & $-$1.57 & 0.14 & 0.05 & $-$1.58 & 0.14 & 0.08\\
271 & 2.82 & 0.18 & $-$1.48 & 0.15 & 0.06 & $-$1.48 & 0.14 & 0.08\\
277 & 2.20 & 0.23 & $-$1.74 & 0.17 & 0.08 & $-$1.76 & 0.17 & 0.10\\
280 & 2.76 & 0.19 & $-$1.50 & 0.15 & 0.06 & $-$1.50 & 0.15 & 0.09\\
281 & 1.46 & 0.20 & $-$2.05 & 0.15 & 0.06 & $-$2.09 & 0.15 & 0.09\\
282 & 2.72 & 0.18 & $-$1.52 & 0.15 & 0.06 & $-$1.52 & 0.14 & 0.08\\
285 & 2.76 & 0.19 & $-$1.50 & 0.15 & 0.06 & $-$1.50 & 0.15 & 0.09\\
293 & 1.95 & 0.18 & $-$1.84 & 0.14 & 0.05 & $-$1.87 & 0.14 & 0.08\\
296 & 1.53 & 0.19 & $-$2.02 & 0.15 & 0.05 & $-$2.06 & 0.15 & 0.09\\
302 & 1.97 & 0.19 & $-$1.83 & 0.15 & 0.05 & $-$1.86 & 0.15 & 0.09\\
304 & 2.01 & 0.17 & $-$1.81 & 0.14 & 0.04 & $-$1.84 & 0.14 & 0.08\\
320 & 2.96 & 0.18 & $-$1.42 & 0.15 & 0.06 & $-$1.41 & 0.14 & 0.08\\
331 & 0.90 & 0.19 & $-$2.28 & 0.15 & 0.04 & $-$2.35 & 0.15 & 0.09\\
333 & 2.89 & 0.21 & $-$1.45 & 0.16 & 0.07 & $-$1.45 & 0.16 & 0.10\\
336 & 1.50 & 0.18 & $-$2.03 & 0.14 & 0.05 & $-$2.08 & 0.14 & 0.08\\
341 & 2.93 & 0.22 & $-$1.43 & 0.17 & 0.08 & $-$1.43 & 0.17 & 0.10\\
351 & 3.01 & 0.19 & $-$1.40 & 0.15 & 0.07 & $-$1.39 & 0.15 & 0.09\\
352 & 3.23 & 0.22 & $-$1.31 & 0.17 & 0.08 & $-$1.29 & 0.17 & 0.10\\
370 & 1.80 & 0.21 & $-$1.90 & 0.16 & 0.07 & $-$1.94 & 0.16 & 0.10\\
377 & 2.22 & 0.20 & $-$1.73 & 0.16 & 0.06 & $-$1.75 & 0.15 & 0.09\\
379 & 2.72 & 0.20 & $-$1.52 & 0.16 & 0.07 & $-$1.52 & 0.15 & 0.09\\
392 & 2.19 & 0.24 & $-$1.74 & 0.18 & 0.09 & $-$1.76 & 0.18 & 0.11\\
395 & 2.24 & 0.19 & $-$1.72 & 0.15 & 0.06 & $-$1.74 & 0.15 & 0.09\\
396 & 2.10 & 0.33 & $-$1.78 & 0.23 & 0.13 & $-$1.80 & 0.23 & 0.15\\
397 & 2.47 & 0.21 & $-$1.62 & 0.16 & 0.07 & $-$1.64 & 0.16 & 0.10\\
402 & 2.28 & 0.21 & $-$1.70 & 0.16 & 0.07 & $-$1.72 & 0.16 & 0.10\\
409 & 1.97 & 0.30 & $-$1.83 & 0.21 & 0.11 & $-$1.86 & 0.21 & 0.14\\
420 & 2.85 & 0.18 & $-$1.46 & 0.15 & 0.06 & $-$1.46 & 0.14 & 0.08\\
422 & 1.32 & 0.21 & $-$2.11 & 0.16 & 0.06 & $-$2.16 & 0.16 & 0.10\\
429 & 2.65 & 0.21 & $-$1.55 & 0.16 & 0.07 & $-$1.55 & 0.16 & 0.10\\
430 & 2.38 & 0.19 & $-$1.66 & 0.15 & 0.06 & $-$1.68 & 0.15 & 0.09\\
432 & 3.84 & 0.35 & $-$1.05 & 0.24 & 0.14 & $-$1.02 & 0.25 & 0.16\\
434 & 2.40 & 0.16 & $-$1.65 & 0.14 & 0.04 & $-$1.67 & 0.13 & 0.07\\
435 & 2.44 & 0.28 & $-$1.64 & 0.20 & 0.11 & $-$1.65 & 0.20 & 0.13\\
441 & 2.72 & 0.23 & $-$1.52 & 0.17 & 0.08 & $-$1.52 & 0.17 & 0.10\\
471 & 3.31 & 0.65 & $-$1.27 & 0.43 & 0.27 & $-$1.26 & 0.44 & 0.29\\
493 & 2.24 & 0.23 & $-$1.72 & 0.17 & 0.08 & $-$1.74 & 0.17 & 0.10\\
495 & 2.45 & 0.21 & $-$1.63 & 0.16 & 0.07 & $-$1.65 & 0.16 & 0.10\\
508 & 2.90 & 0.22 & $-$1.44 & 0.17 & 0.08 & $-$1.44 & 0.17 & 0.10\\
524 & 0.81 & 0.22 & $-$2.32 & 0.16 & 0.06 & $-$2.39 & 0.16 & 0.10\\
526 & 2.76 & 0.22 & $-$1.50 & 0.17 & 0.08 & $-$1.50 & 0.17 & 0.10\\
533 & 2.83 & 0.23 & $-$1.47 & 0.17 & 0.08 & $-$1.47 & 0.17 & 0.10\\
541 & 2.24 & 0.23 & $-$1.72 & 0.17 & 0.08 & $-$1.74 & 0.17 & 0.10\\
549 & 2.70 & 0.24 & $-$1.53 & 0.18 & 0.09 & $-$1.53 & 0.18 & 0.11\\
552 & 2.41 & 0.26 & $-$1.65 & 0.19 & 0.10 & $-$1.66 & 0.19 & 0.12\\
556 & 4.65 & 0.41 & $-$0.71 & 0.28 & 0.17 & $-$0.65 & 0.29 & 0.19\\
558 & 1.86 & 0.21 & $-$1.88 & 0.16 & 0.07 & $-$1.91 & 0.16 & 0.10\\
569 & 2.17 & 0.23 & $-$1.75 & 0.17 & 0.08 & $-$1.77 & 0.17 & 0.10\\
570 & 2.84 & 0.32 & $-$1.47 & 0.23 & 0.13 & $-$1.47 & 0.23 & 0.14\\
573 & 3.34 & 0.21 & $-$1.26 & 0.16 & 0.07 & $-$1.24 & 0.16 & 0.10\\
576 & 3.00 & 0.26 & $-$1.40 & 0.19 & 0.10 & $-$1.40 & 0.19 & 0.12\\
584 & 2.55 & 0.22 & $-$1.59 & 0.17 & 0.08 & $-$1.60 & 0.17 & 0.10\\
588 & 3.12 & 0.26 & $-$1.35 & 0.19 & 0.10 & $-$1.34 & 0.19 & 0.12\\
604 & 2.28 & 0.21 & $-$1.70 & 0.16 & 0.07 & $-$1.72 & 0.16 & 0.10\\
614 & 3.02 & 0.21 & $-$1.39 & 0.16 & 0.08 & $-$1.39 & 0.16 & 0.10\\
623 & 2.06 & 0.21 & $-$1.79 & 0.16 & 0.07 & $-$1.82 & 0.16 & 0.10\\
 \hline  \hline
 \multicolumn{9}{l}{\renewcommand{\baselinestretch}{1.}
\small{$^a$ The \wp values listed here have been scaled to the
work of R97a.}}\\ \label{table-leo2_spec}

 \end{longtable}

%\end{landscape}

\clearpage

\twocolumn
%INSERT FIGURES

\clearpage

\clearpage

\clearpage

\clearpage

\clearpage    

\clearpage

\clearpage

\clearpage

\clearpage

 \clearpage

 \clearpage
\clearpage

\clearpage

\clearpage

\clearpage

\clearpage

\clearpage

\clearpage

\end{document}